\documentclass[entropy,review,accept,pdftex,moreauthors]{Definitions/mdpi} 

\firstpage{1} 
\pubvolume{26}
\issuenum{11}
\articlenumber{947}
\pubyear{2024}
\copyrightyear{2024}
\externaleditor{Academic Editor: Alessandro Pesci}
\datereceived{3 October 2024} 
\daterevised{30 October 2024} 
\dateaccepted{31 October 2024} 
\datepublished{5 November 2024} 
\hreflink{https://doi.org/10.3390/e26110947} 

\Title{Thermodynamics of the Primordial Universe} 

\TitleCitation{Thermodynamics of the Primordial Universe}

\Author{David Silva Pereira $^{1,2}$\orcidC{}, João Ferraz $^{1}$, Francisco S. N. Lobo $^{1,2,}$*\orcidB{} and José Pedro Mimoso $^{1,2,}$*\orcidA{}} 

\AuthorNames{David Silva Pereira, João Ferraz, Francisco S. N. Lobo, José Pedro Mimoso}

\AuthorCitation{Pereira, D.S.; Ferraz, J.;\linebreak Lobo, F.S.N.; Mimoso, J.P.}

\address{$^{1}$ \quad Instituto de Astrof\'{\i}sica e Ci\^{e}ncias do Espa\c{c}o, Faculdade de
Ci\^encias da Universidade de Lisboa, \mbox{Campo Grande, }Edif\'{\i}cio C8,
1749-016~Lisbon,~Portugal; davidsp@alunos.fc.ul.pt~(D.S.P.);\linebreak jferraz@fc.ul.pt~(J.F.) \\
$^{2}$ \quad Departamento de F\'{i}sica, Faculdade de Ci\^{e}ncias da Universidade de Lisboa, Campo Grande, Edif\'{\i}cio C8, 1749-016~Lisbon,~Portugal}

\corres{Correspondence: fslobo@fc.ul.pt~(F.S.N.L.); jpmimoso@fc.ul.pt~(J.P.M.)}

\abstract{This review delves into the pivotal primordial stage of the universe, a period that holds the key to understanding its current state. To fully grasp this epoch, it is essential to consider three fundamental domains of physics: gravity, particle physics, and thermodynamics. The thermal history of the universe recreates the extreme high-energy conditions that are critical for exploring the unification of the fundamental forces, making it a natural laboratory for high-energy physics. This thermal history also offers valuable insights into how the laws of thermodynamics have governed the evolution of the universe's constituents, shaping them into the forms we observe today.
Focusing on the Standard Cosmological Model (SCM) and the Standard Model of Particles (SM), this paper provides an in-depth analysis of thermodynamics in the primordial universe.
The~structure of the study includes an introduction to the SCM and its strong ties to thermodynamic principles. It then explores equilibrium thermodynamics in the context of the expanding universe, followed by a detailed analysis of out-of-equilibrium phenomena that were pivotal in shaping key events during the early stages of the universe's evolution.}

\begin{document}
\section{Introduction}\label{sec:intro}

The primordial stage of the universe represents a crucial epoch that fundamentally shapes our current understanding of cosmology, offering insights into the origins of the universe and its subsequent evolution. To comprehensively describe this stage, it is\linebreak necessary to integrate insights from three fundamental domains of physics: gravity, particle physics, and thermodynamics. These interconnected fields provide a robust framework for understanding the high-energy conditions that dominated the early universe, and that are critical for exploring the unification of the fundamental interactions and the\linebreak development of cosmic structures. The gravitational interaction, as described by the theory of general relativity (GR), plays a foundational role in the dynamics of the early universe.\linebreak GR governs the behavior of spacetime on cosmic scales, providing the mathematical\linebreak framework for understanding how matter and energy influence the curvature of spacetime, and the Standard Cosmological Model (SCM), which is deeply rooted in GR, offers\linebreak a comprehensive explanation of how gravity influenced the expansion and thermal history of the universe. According to the SCM, the gravitational collapse of primordial fluctuations in the early universe led to the formation of the large-scale structures we observe today, such as galaxies, galaxy clusters, and superclusters. These primordial fluctuations are believed to have originated from quantum perturbations during the inflationary period~\cite{Guth:1980zm,Starobinsky:1980te,Linde:1983gd},\linebreak \mbox{ a brief phase} of exponential expansion that occurred just fractions of a second after the \mbox{Big Bang.} As the universe expanded, these fluctuations grew under the influence of gravity, eventually leading to the complex web of structures that populate the cosmos~\cite{Peebles:1980yev,Kolb:1988aj,Kolb:1990vq}.

The role of gravity in the early universe is not only limited to the formation of structures but is also fundamental to the overall evolution of the universe. The interplay between gravity and the thermal energy of the primordial plasma was crucial in determining the rate of expansion and the cooling of the universe. This balance between gravitational forces and thermal pressures set the stage for critical processes like Big Bang nucleosynthesis and recombination. During nucleosynthesis, which occurred within the first few minutes after the Big Bang, the cooling universe allowed protons and neutrons to combine and form the first atomic nuclei, primarily hydrogen, helium, and trace amounts of other light\linebreak elements. This process is pivotal for understanding the elemental composition of the\linebreak universe as predicted by the SCM and confirmed by observations of the cosmic microwave background (CMB) and the abundance of light elements.
Recombination, occurring about 380,000 years after the Big Bang, marks another crucial phase in the evolution of the\linebreak universe. As the universe expanded and cooled further, electrons and protons\linebreak combined to form neutral hydrogen atoms, leading to the decoupling of matter and radiation. This decoupling allowed photons to travel freely through space, giving rise to the CMB radiation, which serves as a fossil record of the early universe. The study of the CMB has provided profound insights into the conditions of the early universe, offering evidence for the SCM and constraints on cosmological parameters such as the Hubble constant, the matter density, and the curvature of the universe~\cite{WMAP:2012fli,Planck:2018vyg}.

The high-energy conditions present in the early universe are fundamentally linked to particle physics, particularly as described by the Standard Model (SM). This model, which has been foundational in our understanding of fundamental particles and their\linebreak interactions, describes the behavior of quarks, leptons, and gauge bosons. During the\linebreak primordial epoch, the universe was in a state of extreme temperature and density,\linebreak existing as a hot, dense plasma of these fundamental particles, and consequently one may consider that this early period of the universe served as a natural laboratory for high-energy physics. The extreme conditions created environments that are unattainable in \mbox{contemporary terrestrial} experiments, providing a unique opportunity to test and explore theoretical frameworks that extend beyond the SM. Among these are the theories of grand unification and supersymmetry, which attempt to unify the electromagnetic, weak, and strong nuclear forces into a single coherent framework~\cite{Georgi:1974sy,Wess:1974tw}. Grand unification theories (GUTs) suggest that at sufficiently high energies, these three fundamental forces merge into a single force, a concept that has profound implications for our understanding of the universe's origin and the fundamental structure of matter. Supersymmetry, on the other hand, proposes a symmetry between fermions and bosons, predicting the existence of superpartner particles for each particle in the SM, which could potentially resolve several outstanding issues, such as the hierarchy problem and the nature of dark matter.

The thermal history of the universe, particularly during critical epochs like the quark--gluon plasma phase and the electroweak phase transition, also offers profound insights into phenomena that cannot be replicated in terrestrial laboratories. The quark--gluon plasma phase, for example, represents a state of matter where quarks and gluons, normally confined within protons and neutrons, exist freely in a hot, dense soup. 
As the universe cooled, it underwent a series of phase transitions that were pivotal in shaping the observable universe. One of the most significant of these transitions was the formation of hadrons from quarks and gluons as the quark--gluon plasma cooled and condensed. This was followed by the decoupling of neutrinos and photons, which occurred as the universe expanded and cooled further. The decoupling of photons, in particular, led to the formation of the above-mentioned CMB, the afterglow of the Big Bang, which provides a snapshot of the universe when it was just 380,000 years old~\cite{Peebles:1970ag,Zeldovich:1969ff,Sunyaev:1970er}. This decoupling event also marks the moment when the universe became transparent to radiation, allowing photons to travel freely through space, carrying with them information about the early universe~\cite{Shuryak:1980tp,Kolb:1988aj}.
Indeed, the early universe, with its extreme conditions and rapid evolution, in the context of the SM offers a unique and invaluable window into the fundamental processes that govern our universe. 

Thermodynamics is central to understanding how the universe evolved from its hot, dense state to the more structured cosmos we observe today. The laws of thermodynamics, particularly the conservation of energy and the increase in entropy, governed the behavior of the primordial plasma and dictated the conditions under which particles could interact and combine into more complex structures 
~\cite{Weinberg:1972kfs,Mukhanov:2005sc}. The adiabatic expansion of the universe, a process described by the first law of thermodynamics, led to the cooling of the plasma, allowing for the synthesis of light elements during Big Bang nucleosynthesis, a process that has been extensively studied and validated through observations of primordial element abundances~\cite{Alpher:1948ve,Steigman:2007xt}.
Moreover, the thermal history of the universe provides a framework for understanding the decoupling of radiation and matter, which gave rise to the CMB. Indeed, the detailed structure of the CMB, including its anisotropies, is a direct consequence of the thermodynamic processes in the early universe and offers critical evidence for the SCM.
These anisotropies, which manifest as tiny temperature fluctuations in the CMB, are imprints of the density fluctuations in the primordial plasma. They provide critical evidence for the SCM, allowing cosmologists to infer key parameters such as the age, composition, and geometry of the universe. The study of these anisotropies has been instrumental in refining our understanding of the early universe, supporting the theory of cosmic inflation and providing insights into the nature of dark matter and dark energy.

Thus, thermodynamics is not just a peripheral aspect but is central to the cosmological\linebreak narrative, providing the framework through which the early universe's evolution\linebreak is understood. From the cooling of the primordial plasma to the formation of the first elements and the decoupling of radiation, thermodynamic principles have shaped the universe's trajectory, leaving imprints that we can observe and study today. These insights are critical for constructing a coherent picture of the universe's origins and its evolution over billions of years.
This paper is devoted to an in-depth study of thermodynamics in the primordial universe, framed within the context of the SCM and SM. The thermal history of the\mbox universe is heavily influenced by the gravitational model adopted, and this study seeks to build a comprehensive description of the primordial universe by integrating \mbox{these models. }

This work is organized in the following manner: In Section \ref{sec:SCM}, we introduce the Standard Model of Cosmology, highlighting its deep connection with the thermodynamic description of the universe.
In Section \ref{sec:thermoPU}, we lay the groundwork for exploring thermodynamics in the context of the primordial epoch.
Section \ref{sec:Out-Of-Equilibrium Phenomena} presents a brief exposition of out-of-equilibrium phenomena, which are crucial for understanding key events in the early universe, such as baryogenesis and the formation of cosmic structures.
In Section  \ref{sec:Inflation}, we provide a brief overview of inflation, and a selection of slow-roll inflationary models that are in agreement with the Planck data. In Section \ref{sec:Brief Thermal History of the Universe}, we present a brief description of the thermal history of the universe.
Finally, in Section \ref{Sec:Conclusion}, we conclude with a comprehensive summary of the main topics presented, synthesizing the insights gained from studying the thermodynamics of the primordial universe and highlighting the ongoing challenges and future directions in this field of research.

\section{Standard Model of Cosmology}\label{sec:SCM}

\subsection{Theoretical Framework}

To built a cohesive description of the primordial universe, a framework that is able to picture the grand scale of the universe is needed.  Currently, the most effective theoretical framework for this purpose is the SCM, which combines GR~\cite{Mimoso:2021sic,Wald:1984rg,Weinberg:1972kfs,dInverno:1992gxs} with the Cosmological Principle~\cite{Perivolaropoulos:2021jda}. The latter asserts that at large scales, the universe appears homogeneous and isotropic. The use of GR establishes the field equations, known as the Einstein field equations, which link the spacetime geometry to non-gravitational fields. These equations can be derived from the principle of least action~\cite{Wald:1984rg,Weinberg:1972kfs}, where the fundamental action consists of the Einstein--Hilbert action and a matter Lagrangian, $\mathcal{L}_m$, and is expressed \mbox{as follows:}
\begin{equation}\label{GR_actiom}
\vspace{-6pt}
    S = \int \mathrm{d}^4 x \sqrt{-g} \left[ \frac{1}{16\pi G} (R- 2\Lambda) + \mathcal{L}_m \right] \ , 
\end{equation}
where $g$ is the determinant of the metric, $G$ is Newton’s gravitational constant, $\Lambda$ is the cosmological constant, and $\sqrt{-g} \, \mathcal{L}_m$ is the Lagrangian density for the matter fields.
By varying this action with respect to the metric degrees of freedom, $g^{\mu \nu}$, one obtains the following field equations
\begin{equation}\label{F_Equations}
    R_{\mu \nu} - \frac{1}{2} R g_{\mu \nu} + \Lambda g_{\mu\nu} = 8 \pi G T_{\mu \nu} \ ,
\end{equation}
where $T_{\mu \nu}$ is the energy--momentum tensor of the matter fields defined as
\begin{equation}\label{T from action varition}
    T_{\mu \nu} = -\frac{2}{\sqrt{-g}} \frac{\delta (\sqrt{-g} \mathcal{L}_m) }{\delta g^{\mu \nu}} \ .
\end{equation}

The Cosmological Principle dictates the appropriate metric to characterize the gravitational field, which is assumed to be the Friedmann--Lemaitre--Roberson--Walker (FLRW) metric, given by
\begin{equation}\label{FLRW}
\mathrm{d}s^2 = -\mathrm{d}t^2 + a^2(t) \left[ \frac{\mathrm{d}r^2}{1-kr^2} + r^2 (\mathrm{d}\theta^2 + \sin^2\theta \, \mathrm{d}^2\phi ) \right] \ ,
\end{equation}
where $a(t)$ represents the scale factor of the universe, while $r$, $\theta$, and $\phi$ are spatial comoving coordinates denoting a radial coordinate and two spherical angular coordinates, respectively. The curvature parameter $k$ can take values of $-1$, $0$, or $+1$, corresponding to a universe with negative, zero (flat), or positive curvature, respectively. The metric signature convention used throughout this work is $(-,+,+,+)$.

Furthermore, we consider an energy--momentum tensor consistent with the symmetries of the FLRW metric, given by~\cite{Weinberg:1972kfs}
\begin{equation}\label{Perfect_fluid}
T_{\mu \nu} = (\rho + p) u_\mu u_\nu + p g_{\mu \nu} \ ,
\end{equation}
where $\rho$ represents the energy density of the universe's matter content, and $p$ is the isotropic pressure. This form of the energy--momentum tensor corresponds to that of a perfect fluid. Substituting this tensor into the Einstein field Equation \eqref{F_Equations}, and considering the FLRW metric (\ref{FLRW}), yields the following dynamical equations
\begin{equation}\label{Friedmann}
    H^2(t) + \frac{k}{a^2(t)} = \frac{8\pi G}{3}\rho + \frac{\Lambda}{3} \,,
\end{equation}
\begin{equation}\label{aceleration}
    \dot{H}(t) + H^2 (t) = - \frac{4\pi G}{3}(\rho + 3p) + \frac{\Lambda}{3} \,,
\end{equation}
where $H \equiv \dot{a}(t)/a(t)$ is the expansion rate of the universe, where the overdot denotes a derivative with respect to cosmic time $t$. 
Equation \eqref{Friedmann} is referred to as the Friedmann equation and is derived from the $00$ component of Einstein's field equations. On the other hand, Equation~\eqref{aceleration} is the Raychaudhuri equation, serving as an expression for the acceleration. It is obtained by subtracting the Friedmann equation from the $i-i$ components of the Einstein field equations. 

Additionally, this set of equations can be used in combination with the conservation equation that results from the fact that the Einstein tensor, $G_{\mu \nu}$, satisfies the contracted Bianchi identities, leading to the conservation of the energy--momentum tensor, i.e., $\nabla_{\mu}T^{\mu \nu} = 0$. The conservation equations reads
\begin{equation}\label{Conservation}
\vspace{-6pt}
\dot{\rho} = -3 H(t)(\rho + p) \ ,
\end{equation}
which stems from the conservation of energy and is inherent in the structure of the gravitational field equations. To complete the system of equations governing the evolution of the scale factor, it is necessary to introduce an appropriate barotropic equation of state, i.e., $p=w \rho$. For all components, the simplest equation of state is provided by
\begin{equation}\label{baritropic PFluid}
\sum_i p_i = \sum_i (\gamma_i-1) \rho_i \ ,
\end{equation}
where $\gamma_i$ is a constant parameter, and $\gamma_i-1$ represents the speed of sound, $v^2_s$, in the fluid $i$ characterizing the $i$-th component of the universe's matter content. This equation of state facilitates the integration of the energy conservation Equation \eqref{Conservation}, leading to
\begin{equation}\label{densidade}
    \rho (t) = \sum_{i} \rho_{i,0} \, \left(\frac{a(t)}{a_0}\right)^{-3\gamma_i} = \sum_i \rho_{i,0} (1+z)^{-3\gamma_i} \ ,
\end{equation}
where $\rho(t) \equiv \sum_i \rho_i(t)$ represents the total energy density of the universe at a given time $t$ and  $\rho_{i,0} \equiv \rho_i(t_0)$ and $a_0 \equiv a(t_0) $ are constants of integration which, without loss of generality, can be respectively established by the values of each component $i$ of the matter content of the universe and the value of the scale factor at the present day, $t=t_0$.

\subsection{Evolution of the Universe from the Friedmann Equations}\label{subsec:Evolution of the Universe from the Friedmann equations}

To analyze the time evolution of the universe, or the dynamic behavior of the scale factor,  the universe's history is typically divided into several cosmological epochs, each dominated by a single fluid with a constant parameter $\gamma_i$. Key epochs of interest include the radiation fluid ($i=R$), where $p=\rho/3$ ($\gamma_R=4/3$); incoherent matter ($i=M$), where $p=0$ ($\gamma_M=1$); and vacuum energy ($i=V$), where $p=-\rho$ ($\gamma_V=0$), which is equivalent to a cosmological constant. Additionally, stiff matter fluid ($i=S$), characterized by $p=\rho$ ($\gamma_S=2$), may also be relevant.

From Equation \eqref{densidade}, the density profile for each epoch is $\rho(t)\simeq\rho_i(t)\propto a(t)^{-3\gamma_i}$. Substituting this profile into the Friedmann Equation \eqref{Friedmann} shows that for $\gamma_i>2/3$, the curvature term, $k^2/a^2$, dominates only at later times, assuming a cosmological constant does not dominate earlier. Thus, for early universe modeling, assuming a flat model ($k=0$) is reasonable unless otherwise specified. This simplifies the Friedmann Equation \eqref{Friedmann} and allows for integration, yielding the cosmological solutions
\begin{align}
a(t)\propto t^{\frac{2}{3\gamma_i}}, \qquad H(t)= \frac{2}{3\gamma_i}\frac{1}{t}, \qquad &\textrm{if}~\gamma_i\neq 0,\label{expansion rate non vacuum epochs}
	\\
a(t)\propto e^{\sqrt{\frac{\Lambda}{3}}t}, \qquad H(t)= \sqrt{\frac{\Lambda}{3}}, \qquad &\textrm{if}~\gamma_i=0\,.\label{deSitterGR}
\end{align}

These solutions indicate that the early universe was radiation-dominated, the adolescent universe was matter-dominated, and without vacuum energy, and the late universe would remain matter-dominated. However, current observations suggest otherwisem as the accelerated expansion rate of the universe cannot be achieved in the current framework with only matter-dominated content~\cite{SupernovaSearchTeam:1998fmf,SupernovaCosmologyProject:1998vns}. Furthermore, if the universe experienced an initial period of inflation, there was a very early epoch dominated by vacuum energy, leading to an exponential expansion characterized by the cosmological solution \eqref{deSitterGR}.

Additionally, the Friedmann equation can be expressed in a more convenient form through the introduction of two dimensionless parameters: the critical density, determined by setting $\Lambda = k = 0$ in the aforementioned equation, defined as
\begin{equation}\label{critical density}
\rho_c = \frac{3H^2(t)}{8 \pi G} \ ,
\end{equation}
and the dimensionless density parameter, denoted as $\Omega(t)$, representing the ratio of the energy density $\rho$ to the critical density ($\rho_{c}$) at a given time
\begin{equation}\label{density parameter}
\Omega(t) \equiv \frac{\rho(t)}{\rho_c(t)} \ .
\end{equation}

With these definitions, Equation~\eqref{Friedmann} can be reformulated as
\begin{equation}\label{Omega}
\frac{k}{a^2H^2(t)} = \Omega(t) - 1 \ .
\end{equation}

This establishes a correlation between the sign of $k$ and the sign of $\Omega - 1$.\linebreak Specifically, for $k = 0$ (indicating a flat model), $\Omega = 1$; for $k = 1$ (indicating a closed model), $\Omega > 1$; and for $k=-1$ (indicating an open model), $\Omega<1$. 

Therefore, using Equation~\eqref{densidade} and taking into account that the present day value of the critical density parameter is given by $\rho_{c,0} = 3 H^2_0/(8 \pi G)$, one can write the density parameter in terms of the redshift $z$ as follows
\begin{equation}\label{density_parameter_redshift}
    \Omega(z) = \frac{H^2_0}{H^2(z)} \left[\sum_i \Omega_{i,0} (1+z)^{3\gamma} + \Omega_{\Lambda} \right] \ ,
\end{equation}
where $\Omega_{\Lambda} \equiv \Lambda/(3H^2_0)$ and $\Omega_{i,0} \equiv \rho_{i,0}/\rho_{c,0}$ are, respectively, the contributions from the\linebreak cosmological constant and from the $i$-th fluid component for the present day\linebreak density parameter. 

Moreover, the density parameter can be expressed in terms of \emph{partial density parameters} assigned to a specific component or species $i$ as $\Omega_0=\sum_i\Omega_{i,0}=\sum\rho_{i,0}/\rho_{c,0}$, where $i=R$ denotes relativistic species and $i=M$ represents non-relativistic species. By integrating the equation for the conservation of energy \eqref{Conservation}, using the barotropic Equation \eqref{baritropic PFluid}, the Friedmann equation can ultimately be formulated as
\begin{equation}
H^2(z)+\frac{k^2}{a^2(z)}=H_0^2\left[\sum_i\Omega_{i,0}(1+z)^{3\gamma_i}+\Omega_{\rm{v}}\right]\label{friedmann3}\,.
\end{equation}

\section{Equilibrium Thermodynamics in the Expanding Universe}\label{sec:thermoPU}

To comprehensively elucidate the history of the early universe, it is imperative to track the thermal evolution of its primary constituents. The early universe is known to have been in a state close to thermal equilibrium, a conclusion drawn from the study of the CMB radiation. Since its discovery in 1965~\cite{Penzias:1965wn}, the CMB has been extensively measured, with the most recent and precise observations provided by the Planck satellite~\cite{Planck:2018vyg}.\linebreak The CMB is one of the most important observables of the primordial universe. After accounting for perturbation effects, the CMB closely matches the spectrum of a black body. This near-perfect black body spectrum persisted until the decoupling of neutrinos.

In a system at thermal equilibrium, as was the case for much of the primordial universe, statistical mechanics can be utilized to calculate key quantities such as energy density, pressure, and entropy density. These quantities depend on the particle number density at any given time, with the contributions from different particles being primarily influenced by their properties, particularly mass and degeneracy. For much of the early universe, equilibrium thermodynamics provides a good approximation for describing these conditions. This section will cover the essential principles needed for such a description. For a more detailed discussion, the following references are recommended, as this review draws heavily from them, especially from~\cite{Kolb:1990vq,Bernstein:1988bw, Lindquist:1966igj, Dodelson:2003ft, Baumann:2022mni, Peter:2013avv, Rubakov:2017xzr, Husdal:2016haj}. Additionally, several sections related to cosmology in the \textit{Review of Particle Physics}~\cite{ParticleDataGroup:2012pjm, ParticleDataGroup:2018ovx, ParticleDataGroup:2020ssz, ParticleDataGroup:2022pth} have also been consulted.

\subsection{Kinetic and Chemical Equilibrium}\label{subsec:Kinetic_EQ}

In the realm of thermodynamics, when an isolated system persists over a considerable period, it reaches a state of thermal equilibrium. At this point, all its observable features, like the distribution of particles, settle into their most likely configurations. This includes various macroscopic properties such as particle density (represented by $n$), energy \linebreak density ($\rho$), pressure ($p$), and entropy density ($s$). These quantities are expressed as integrals using a distribution function denoted as $f(x^a,p^a)$. Taking in account the Cosmological\linebreak Principle that states that the universe, at large scales, is spatially homogeneous and isotropic, one is led to assume that the phase space density must be isotropic and homogeneous. Consequently, the distribution function can be simplified to $f(x^a,p^a) = f(|\vec{\textbf{p}}|,t) = f(E, t)$, where $E = \sqrt{\vec{\textbf{p}}^2 + m^2}$ represents the relativistic energy~\cite{Kolb:1990vq,Bernstein:1988bw,Lindquist:1966igj}. Thus, disregarding its explicit dependence on time (which will be revealed through its connection to temperature), the phase space distribution function for a specific particle species $i$ in kinetic equilibrium, within the ideal gas approximation, takes on the form of either the Fermi--Dirac (FD) or Bose--Einstein (BE) distributions~\cite{Kolb:1990vq,Bernstein:1988bw,Lindquist:1966igj}
\begin{equation}\label{D_function}
    f_i(\vec{p}) = \left[ \exp\left(\frac{E_i(\vec{p}) - \mu_i}{T}\right)\pm 1 \right]^{-1} \ ,
\end{equation}
where $E_i$ denotes the energy associated with the particle species denoted by $i$, and $m_i$\linebreak represents its respective rest mass. Furthermore, $u_i$ is the chemical potential of the particle species, where the superscript $(+)$ indicates fermions adhering to Fermi--Dirac (FD) statistics, and the superscript $(-)$ designates bosons following Bose--Einstein (BE) statistics.\linebreak Suppressing the term $\pm 1$ leads to an expression that aligns with the\linebreak classical and distinguishable particles approximation characteristic of Maxwell--Boltzmann (MB) statistics. While this approximation does not yield an exact distribution function, it holds significance, particularly when a degenerate Fermi species ($u_i \gtrsim T_i$) or a Bose condensate is absent. In such cases, the adoption of MB statistics introduces only a modest quantitative deviation from exact statistics, as noted in~\cite{Kolb:1990vq}. Notably, this deviation is often confined to less than 10\%~\cite{Mather:1990tfx}. An advantageous aspect of this approximation emerges in scenarios involving non-relativistic species of particles, specifically when $m_i \gg T_i$, and $m_i \gg T + \mu_i$, leading to the exactness of MB statistics.

To complement the previous description, one has to explore the dynamics of the chemical potential. This quantity signifies the infinitesimal alteration in energy resulting from the inclusion of a new particle of a specific type, albeit being undisclosed. This essential thermodynamic attribute, applicable in scenarios involving a variable number of particles, is encapsulated by the fundamental relation that assumes the following form~\cite{Kolb:1990vq}
\begin{equation}\label{Entropia}
    dE =  T dS - p dV + \sum_i \mu_i dN_i \ ,
\end{equation}
where S is the entropy, $V$ is the volume, and $N_i$ is the particle number of the species $i$.\linebreak This enables the establishment of a condition for interactions commonly observed in an equilibrium gas: $\sum_i \mu_i dN_i = 0$. The significance of this condition lies in its role in maintaining the equilibrium state. A deviation from this condition would necessitate a change in the number of particles for different species, effectively lowering the free energy expressed as $F = E - T S$, where $E$ denotes the internal energy, and under conditions of constant temperature and volume, one has the following differential relation:
\begin{equation}
    dF = \sum_i \mu_i dN_i \ .
\end{equation}

Hence, to sustain equilibrium, encompassing both kinetic and chemical equilibrium, the particle distribution functions in the equilibrium gas have to not only conform to their customary thermal expressions as outlined in Equation \eqref{D_function} but also exhibit an\linebreak interrelation between the chemical potentials of distinct particle species engaged\linebreak in interactions. Specifically, under conditions of chemical equilibrium, the chemical \linebreak potential $\mu_i$ becomes additively conserved across all reactions. Consequently, if \linebreak equilibrium persists in a reaction such as $a_1 + a_2 + ... \leftrightarrow b_1 + b_2 + ...$, the chemical \linebreak potentials of the involved particles are related by the following expression
\begin{equation} \label{Chemical equality}
    \sum_k \mu_{ak} = \sum_{k} \mu_{b_k} \ .
\end{equation}

Nevertheless, concerning radiation, photons exhibit the ability to be emitted or\linebreak absorbed in arbitrary reactions and, for any charged particle, there exists an inelastic\linebreak scattering reaction given by $i + i \to i + i + \gamma$, where $i$ represents a particle species.\linebreak These reactions imply that under conditions of chemical equilibrium, the relation\linebreak $\mu_i + \mu_i = \mu_i + \mu_i + \mu_\gamma$ holds. Consequently, in a state of complete equilibrium, the photon manifests a chemical potential of zero. When scrutinizing reactions involving the annihilation of particles and antiparticles, such as $i + \overline{i} \leftrightarrow 2\gamma, 3\gamma...$, the equilibrium state dictates a correlation between the chemical potentials of particles and antiparticles, characterized by
\begin{equation}
    \mu_{i} + \mu_{\overline{i}} = 0 \ ,
\end{equation}
this leads to the inference that the chemical potentials of particles and their\linebreak corresponding antiparticles possess equal magnitudes but opposite signs. Furthermore, in specific scenarios where particle species exhibit self-conjugation ($i = \overline{i}$) or a symmetry prevails between particles and antiparticles ($n_i = n_{\overline{i}}$), maintaining chemical equilibrium in annihilations dictates that $\mu_{i} = \mu_{\overline{i}} = 0$. Consequently, it becomes evident that under such circumstances, the distribution functions are solely determined by a single\linebreak parameter, namely, the temperature. In essence, the distribution function transitions from being expressed as $f(x^a,p^a)$ to the simplified form $f(T)$.

To further build a more concise description, it proves beneficial to introduce chemical potentials associated with conserved quantum numbers $Q(a)$. For a given particle species $i$, the chemical potential is expressed as
\begin{equation}\label{Chemical Quantum}
\vspace{-6pt}
\mu_i = \sum_k \mu_k Q_i^{(k)} \ ,
\end{equation}
where $Q_i^{(k)}$ signifies the quantum numbers carried by particle $i$. Importantly, these\linebreak quantum numbers are assumed to be independent of each other and collectively form a \mbox{comprehensive set of} conserved quantum numbers as outlined by Dolgov~\cite{A.D. Dolgov}.\linebreak If these quantum numbers remain conserved across all reactions, the relation presented in \linebreak Equation \eqref{Chemical equality} automatically holds. Consequently, all chemical potentials can be expressed in terms of the chemical potentials associated with these conserved quantities. Thus, the number of chemical potentials corresponds to the number of independent conserved quantum numbers. In instances where the charge densities $n_{Q^{(k)}}$ are known, the determination of the chemical potentials for the charges $Q^{(k)}$ becomes feasible. In this context, the number density of each particle species $n_i$ can be expressed as a function of $\mu_i$, resulting in a system of equations
\begin{equation}
\sum_i Q_{i}^{(k)} n_i = n_{Q^{(k)}},
\end{equation}
and in conjunction with Equation \eqref{Chemical Quantum}, this system comprehensively determines all $\mu_i$ in terms of $n_{Q_i}$~\cite{A.D. Dolgov}.

Considering the current SM, there are five independent conserved charges present: electric charge, baryon number, electron--lepton number, muon--lepton number, and\linebreak tau--lepton number. These lead to five corresponding independent chemical potentials~\cite{Weinberg:1972kfs}. These chemical potentials are determined by the respective number densities of the\linebreak conserved quantities. Within this topic, from all the elementary particles, the chemical potential of neutrinos is a delicate subject, and we endorse~\cite{Lesgourgues:1999wu,Mangano:2010ei} for more details.

\subsection{Gases of Free Particles and Interactions in an Expanding Universe}\label{subsec:Gases of free particles and interactions in an Expanding Universe}

Given the preceding discussion, it is crucial to be cautious when applying\linebreak equilibrium thermodynamics to dynamic situations like the expanding universe.\linebreak In such cases, both the temperature ($T$) and chemical potential ($\mu$) evolve to maintain energy continuity and particle number conservation. Importantly, an expanding universe is inherently non-equilibrium, and as a result, the distribution function may deviate from standard equilibrium forms, and its time-dependent behavior may require\linebreak explicit consideration.

For example, in gases consisting of stable, non-interacting particles, namely, those that do not decay or interact significantly with their surroundings, i.e., freely propagating particles, their number within a proper volume $\mathrm{d}^3\mathbf{x}$ in the interval $\mathrm{d}^3\mathbf{p}$ is a conserved quantity~\cite{Kolb:1990vq,Bernstein:1988bw}, and can be expressed as  
\begin{equation}
dN_i=\frac{g_i}{(2\pi)^3}f_i(\mathbf{x},~\mathbf{p},~t)\,\mathrm{d}^3\mathbf{x}\mathrm{d}^3\mathbf{p}=\textrm{Cte}\, ,\label{constant dn in stable particles}
\end{equation}
where $g_i$ is the internal degrees of freedom (number of spin states) associated with the particle $i$~\cite{Kolb:1990vq,Rubakov:2017xzr,Husdal:2016haj}. Assuming the SCM, gases containing freely moving particles are expected to exhibit both homogeneity and isotropy. This means that their distribution functions become independent of spatial coordinates and depend only on momenta and time, as noted before. Furthermore, according to the FLRW geodesic equation of motion, governing the paths of these free particles, the physical three-momentum decreases with the scale factor as $(|\mathbf{p}| \propto a^{-1})$, and the proper volume spatial element behaves as $(d^{3}\mathbf{x} \propto a^{3})$.

Now, examining Equation~\eqref{constant dn in stable particles}, it becomes evident that if the distribution function is known at a specific time $(t = t_d)$ and expressed as $f_i(|\mathbf{p}_d|, t_d) \equiv f_{id}(|\mathbf{p}_d|)$, then at a later time $(t > t_d)$, the distribution function follows the relation $f_{i}(|\mathbf{p}|, t)d^{3}\mathbf{x}\mathrm{d}^3\mathbf{p} = f_{id}(|\mathbf{p}_d|)\mathrm{d}^3\mathbf{x}_d\mathrm{d}^3\mathbf{p}_d = f_{id}(a|\mathbf{p}|/a_d)d^{3}\mathbf{x}\mathrm{d}^3\mathbf{p}$. In simpler terms, this implies that the distribution function is entirely determined by the redshifted momentum. Therefore, if a particular stable particle species is in equilibrium and suddenly decouples from its surrounding medium at a time $(t = t_d)$, its distribution function before the decoupling $(t \leq t_d)$ adheres to its usual thermal form, while after the decoupling $(t > t_d)$, it is simply given by $f^{eq}_i(a|\mathbf{p}|/a_d)$.

When accounting for interactions, it is essential to determine whether the particles are coupled or decoupled. A preliminary step in this analysis involves comparing the interaction rate, $\Gamma_{\textrm{int}}$, associated with the relevant processes of the species, to the expansion rate of the universe:
\vspace{-6pt}
\begin{align}
\Gamma_{\textrm{int}}&>H, \qquad \textrm{coupled}\,,\label{coupled}\\
\Gamma_{\textrm{int}}&<H, \qquad \textrm{decoupled}\,.\label{decoupled}
\end{align}

Although true equilibrium is difficult to achieve in an expanding universe, the gradual nature of expansion generally allows the particle composition to approach a state close to local equilibrium. Due to the universe's homogeneity, local thermodynamic quantities effectively reflect global values. Therefore, if particle interactions are much faster than the universe's expansion rate, equilibrium-like distributions of particles can emerge on the timescale of expansion. For a particular species of particles, this implies that it will pass through successive states resembling equilibrium whenever the condition (\ref{coupled}) is met.\linebreak In essence, the thermal history of the universe can be seen as a competition between the rate of particle interactions and the rate of cosmic expansion. However, when $\Gamma_{\textrm{int}} < H$, this does not necessarily mean that equilibrium will break down. For such a departure from equilibrium to occur, the rate of a key reaction essential for maintaining equilibrium must remain lower than the Hubble parameter $H$.

If a particular stable and massless particle species $i$ with a vanishing chemical potential ($\mu_i = 0$) is in equilibrium at a time $t < t_d$, and then it abruptly decouples at time $t_d$, the distribution function of the particle species at a later time will be simply determined by 
\begin{equation}
f_i(\vert\mathbf{p}\vert,~t)=f_i^{eq}\left(\frac{a(t)\vert\mathbf{p}(t)\vert}{a_d}\right)=\left[\exp\left(\frac{\vert\mathbf{p}(t)\vert a(t)}{T_d a_d}\right)\pm 1\right]^{-1}=f_i^{eq}\left(\frac{\vert\mathbf{p}\vert}{T_i}\right)\, . \label{massless decoupled species1}
\end{equation} 

Here, $T_d=T(t_d)$ and $a_d=a(t_d)$ represent the temperature and the scale factor at the moment of decoupling, respectively. The temperature of the decoupled species at any subsequent time $t$ is given by
\begin{equation}
\vspace{-6pt}
T_i(t)=\frac{a_d}{a(t)}T_d\,.
\label{temperature of massless decoupled species}
\end{equation}

Despite the fact that the decoupled particle species may not be in thermal equilibrium, its distribution function possesses the same shape as the equilibrium distribution function, $f^{eq}$, for massless particle species. However, the effective temperature is always decreasing as $T_i\propto a^{-1}$. In particular, for observed decoupled photons of the CMB with approximately zero or negligibly small chemical potential ($\vert\mu_{\gamma}/T_{\gamma}\vert<10^{-4}$)~\cite{A.D. Dolgov}, it can be inferred that if the universe was ever in equilibrium, the photon distribution has consistently been Planckian (given by Equation (\ref{massless decoupled species1}) with the $-$ sign). Thus, the photon number density, energy density, and pressure can be straightforwardly computed and have consistently been given by 
\begin{eqnarray}
n_{\gamma}&=&\frac{g_{\gamma}}{(2\pi)^3}\int f^{Pl}_{\gamma}(\vert\mathbf{p}\vert)\mathrm{d}^{3}\mathbf{p}
	=\frac{g_{\gamma}}{2\pi^2}\int\frac{E^2_{\gamma}}{e^{E_{\gamma}/T_{\gamma}}-1}\mathrm{d}E_{\gamma}=\frac{2\zeta(3)}{\pi^2}T_{\gamma}^3 \ ,
	\label{photon number density}\\
\rho_{\gamma}&=&\frac{g_{\gamma}}{(2\pi)^3}\int E_{\gamma} f^{Pl}_{\gamma}(\vert\mathbf{p}\vert)\mathrm{d}^{3}\mathbf{p}
	=\frac{g_{\gamma}}{2\pi^2}\int\frac{E^3_{\gamma}}{e^{E_{\gamma}/T_{\gamma}}-1}\mathrm{d}E_{\gamma}=\frac{\pi^2}{15}T_{\gamma}^3 \ ,
	\label{photon energy density}\\
p_{\gamma}&=&\frac{g_{\gamma}}{(2\pi)^3}\int \frac{\vert\vec{\mathbf{p}}\vert^2}{3E_{\gamma}} f^{Pl}_{\gamma}(\vert\mathbf{p}\vert)\mathrm{d}^{3}\mathbf{p}
	=\frac{g_{\gamma}}{6\pi^2}\int\frac{E^3_{\gamma}}{e^{E_{\gamma}/T_{\gamma}}-1}\mathrm{d}E_{\gamma}=\frac{1}{3}\rho^{eq}_{\gamma}=\frac{\pi^2}{45}T_{\gamma}^3\label{photon pressure}\,,
\end{eqnarray}
respectively, where $g_\gamma = 2$ accounts for the two photon polarization states and an integration over all angles is performed so that $\mathrm{d}^3 \mathbf{p}=4\pi \vert\vec{\mathbf{p}}\vert^2 d\vert\vec{\mathbf{p}}\vert$ with $E_{\gamma}=\vert\vec{\mathbf{p}}\vert$. In the context of maintaining thermal equilibrium with the surrounding medium, photons exhibit a shared temperature with the equilibrium thermal background ($T_{\gamma}=T$). However, upon decoupling, the temperature of the photons undergoes a continual decrease with the scale factor, expressed as $T_{\gamma}=a_d/a(t)T_{d}$. The number density of any other potentially decoupled particle species will also always be given by
\begin{equation}
n_i=\frac{g_i}{(2\pi)^3}\int \left[\exp\left(\frac{\vert\mathbf{p}\vert a}{T_d a_d}\right)\pm 1\right]^{-1}\mathrm{d}^3\mathbf{p}=g^n_{\textrm{eff}_i}\left(\frac{\zeta(3)T_\mathrm{d}^3}{\pi^2}\right)\,\left(\frac{a_d}{a}\right)^3\,,\label{number density of decoupled massless particle species 1a}
\end{equation}
where $g^n_{\textrm{eff}_i}=g_{\textrm{eff}_i}=g_i$ or $g^n_{\textrm{eff}_i}=3/4g_i$ depending on whether the particle species $i$ is a boson or a fermion respectively. Under the condition that these particle species are in equilibrium with photons prior to their decoupling, their number density at any given time remains consistently comparable to the photon number density. Consequently, each of these decoupled species persists as a relic background in the universe. Therefore, the temperature at which decoupling occurs is often called the ``freezing temperature'', denoted as $T_f$. Its rough estimation is facilitated through a relationship commonly known as the ``freezing relation'' 
\begin{equation}
\Gamma_{\textrm{int}}(T_f)=H(T_f)\, .\label{freezing relation1}
\end{equation}

Alternatively, if a specific particle species, denoted as $i$, undergoes decoupling at a time $t_d$ when it has already transitioned to a non-relativistic state ($T_d \ll m_i$), and the condition $m_i - \mu_i \gg T$ is satisfied, the distribution function of this particle species reverts to the MB distribution function at the moment of decoupling. In this scenario, the kinetic energy is notably lower than the rest mass $m_i$, enabling the approximation $E_i = m_i + p^2/(2m_i)$. As a result, at later times ($t > t_d$), the distribution of the particle species can be precisely characterized by
\begin{equation}
f_i(\vert\vec{\mathbf{p}}\vert, t)=f^{MB}_i\left(\frac{a(t)}{a}p(t)\right)=\frac{1}{(2\pi)^3}\exp\left(-\frac{m_i-\mu_{iD}}{T_d}\right)\exp\left(\frac{\vert\vec{\mathbf{p}}\vert}{2m_i^2}\frac{a(t)^2}{a_d^2T_d^2}\right)\,,
\end{equation}
where $\mu_{iD}$ is the species chemical potential at the time of the decoupling. Therefore, post-decoupling, the distribution function of the particle species retains its equilibrium form, albeit with an effective chemical potential denoted as $\mu_i(t)$, and temperature denoted as $T_i$. These are expressed as follows:
\begin{eqnarray}
\mu_i(t)=m_i+\left(\frac{\mu_{iD}-m_i}{T_d}\right)T_i(t)\,,  \qquad
T_i(t)=\left(\frac{a_d}{a(t)}\right)^2T_d\,.
\label{temperature of non relativisitic decoupled species}
\end{eqnarray}

This phenomenon is distinctly indicative of an effective temperature that diminishes at a rate surpassing that observed in the instance of decoupled massless particle species. Particles decoupling in a non-relativistic state are often viable candidates for what is referred to as ``cold dark matter''. 
In this context, they persist indefinitely in the universe, manifesting as a cold relic background.

However, if a particular particle species $i$ possesses a small but non-zero mass and is\linebreak relativistic at the point of decoupling ($m_i\ll T_d$), its distribution function becomes ``frozen'' in the configuration of a massless particle species distribution function.\linebreak Consequently, its spectrum remains thermal, with the effective temperature declining with the scale factor $a^{-1}$ as long as $T\gg m_i$. However, once the thermal bath's temperature drops below the particle's mass, the distribution function and number density adhere to the\linebreak frozen-in-form characteristic of relativistic particles. Despite this, the equilibrium\linebreak distribution function assumes the MB distribution form, and the energy density\linebreak corresponds to that of a non-relativistic particle \textls[-5]{species, i.e., $\rho\simeq mn$. As a result, the distribution function no longer aligns with an equilibrium distribution, marked by $T\propto a^{-1}$. This scenario characterizes what is known as ``hot dark matter'' and \mbox{``warm dark matter''~\cite{Menci:2016eui,Hannestad:2010yi,Primack:1983xj,Bond:1983xpx}}.} In these instances, the particle species decouples in a relativistic state, maintaining massless distribution functions in momenta.

In the absence of interactions, particles decoupling at a semi-relativistic temperature ($T_d\sim m$) also deviate from maintaining an equilibrium distribution. While the provided description remains highly accurate in many scenarios, it is crucial to emphasize that for a comprehensive understanding of the evolution of particle phase space distributions, particularly in the case of massive particle species, the Boltzmann equations should be rigorously solved.

\subsection{The Matter Content of the Universe}\label{subsec:The matter content of the Universe}

When equilibrium exists among the primary components of the universe's matter content, the total thermodynamic quantities, such as the total number density, energy density, and pressure, can be accurately estimated by considering only the contributions from species that are in equilibrium. In these situations, calculating these quantities becomes straightforward as previously discussed.
If the universe was ever in a state of equilibrium, it is reasonable to expect that, in many cases, even when equilibrium among the main components is no longer maintained, the material composition of the universe can still be effectively described by including contributions from particle species that follow thermal distribution functions, albeit with temperatures different from the equilibrium temperature of the cosmic medium, denoted as $T$. Incorporating these contributions into the overall equilibrium quantities is often beneficial and simple to do. Thus, if the\linebreak non-equilibrium components can be described by an equilibrium distribution function, the total number density, energy density, and pressure can be expressed as follows:
\begin{align}\label{number density}
n=\sum_i\frac{g_i}{(2\pi)^3}\int f_i(\vec{\mathbf{p}})\,\mathrm{d}^3p=&\sum_i\frac{g_i}{2\pi^2}\int_{m_i}^{\infty}\frac{\sqrt{E_i^2-m_i^2} E_i}{\exp[(E_i-\mu_i)/T_i]\pm1}\,\mathrm{d}E_i 
\nonumber \\
=&\frac{T^3}{2\pi^2}\mathcal{N}_{\ast}\left(\left\{\frac{T_i}{T}\right\},\,\left\{\frac{m_i}{T_i}\right\},\,\left\{\frac{\mu_i}{T_i}\right\}\right)\, ,
\end{align} 
\begin{align}\label{total energy density}
\rho=\sum_i\frac{g_i}{(2\pi)^3}\int E_i(\vec{\mathbf{p}}) f_i(\vec{\mathbf{p}})\,\mathrm{d}^3p=&\sum_i\frac{g_i}{2\pi^2}\int_{m_i}^{\infty}\frac{\sqrt{E_i^2-m_i^2} E_i^2}{\exp[(E_i-\mu_i)/T_i]\pm1}\,\mathrm{d}E_i 
\nonumber \\
=&\frac{\pi^2T^4}{30}\mathcal{E}_{\ast}\left(\left\{\frac{T_i}{T}\right\},\,\left\{\frac{m_i}{T_i}\right\},\,\left\{\frac{\mu_i}{T_i}\right\}\right)\, ,
\end{align} 
\begin{align}\label{total pressure}
p=\sum_i\frac{g_i}{(2\pi)^3}\int\frac{\vert\vec{\mathbf{p}}\vert^2 f_i(\vec{\mathbf{p}})}{3E_i(\vec{\mathbf{p}})}\,\mathrm{d}^3p=&\sum_i\frac{g_i}{6\pi^2}\int_{m_i}^{\infty}\frac{(E_i^2-m_i^2)^{3/2}}{\exp[(E_i-\mu_i)/T_i]\pm1}\,\mathrm{d}E_i 
\nonumber \\
=&\frac{\pi^2T^4}{90}\mathcal{P}_{\ast}\left(\left\{\frac{T_i}{T}\right\},\,\left\{\frac{m_i}{T_i}\right\},\,\left\{\frac{\mu_i}{T_i}\right\}\right)\, ,
\end{align}
\textls[-10]{where the second relation is obtained performing integration over angles using $EdE=\vert\vec{\mathbf{p}}\vert\, d \vert\vec{\mathbf{p}}\vert$ and $\mathcal{E}_{\ast}(\{x_i\},~\{z_i\}~\{y_i\})=\sum_i\mathcal{E}_i^{\pm}(x_i,~z_i,~y_i)$, $\mathcal{N}_{\ast}(\{x_i\},~\{z_i\},~\{y_i\})=\sum_i\mathcal{N}_i^{\pm}(x_i,~z_i,~y_i)$} and $\mathcal{P}_{\ast}(\{x_i\},~\{z_i\},~\{y_i\})=\sum_i\mathcal{P}_i^{\pm}(x_i,~z_i,~y_i)$ can be respectively viewed as a convenient parametrization of the \emph{effective degrees of freedom}~\cite{Kolb:1990vq,Rubakov:2017xzr,Husdal:2016haj} in number density, energy density, and pressure of a genuine BE gas in thermal equilibrium, namely, $\rho=\mathcal{E}_{\ast}\rho_{\gamma}/g_{\gamma}$, $n=\mathcal{N}_{\ast}n_{\gamma}/\zeta(3)g_{\gamma}$ and $p=\mathcal{P}_{\ast}p_{\gamma}/g_{\gamma}$, where
\begin{eqnarray}
\mathcal{N}_i^{\pm}\left(\frac{T_i}{T}~,z_i=\frac{m_i}{T_i},~y_i=\frac{\mu_i}{T_i}\right)
&\equiv & g_i\,\left(\frac{T_i}{T}\right)^3\int_{z_i}^{\infty}\frac{\sqrt{u^2-z_i^2}u}{\exp(u-y_i)\pm 1}\,\mathrm{d}u
	\label{mathcal NC}\ ,\\
\mathcal{E}_i^{\pm}\left(\frac{T_i}{T}~,z_i=\frac{m_i}{T_i},~y_i=\frac{\mu_i}{T_i}\right)
&\equiv &\frac{15}{\pi^4}g_i\,\left(\frac{T_i}{T}\right)^4\int_{z_i}^{\infty}\frac{\sqrt{u^2-z_i^2}u^2}{\exp(u-y_i)\pm 1}\,\mathrm{d}u
	\label{mathcal EC}\ , \\
\mathcal{P}_i^{\pm}\left(\frac{T_i}{T}~,z_i=\frac{m_i}{T_i},~y_i=\frac{\mu_i}{T_i}\right)
&\equiv &\frac{15}{\pi^4}g_i\,\left(\frac{T_i}{T}\right)^4\int_{z_i}^{\infty}\frac{(u^2-z_i^2)^{3/2}}{\exp(u-y_i)\pm 1}\,\mathrm{d}u
	\label{mathcal PC}\, \ ,
\end{eqnarray}
where $u=E/T$, and the expressions illustrate the contributions of each species to the overall degrees of freedom in the aforementioned thermodynamic quantities. 

As previously discussed, the statistical approach used here aims to link the key thermodynamic quantities—the number density, total energy density, and total pressure—with the temperature of the medium involved. To accomplish this, we express all relevant quantities in terms of the photon temperature, which represents the equilibrium temperature of the medium. The parameters $\mathcal{N}\ast, \mathcal{E}\ast, \mathcal{P}_\ast$ are introduced to simplify the analysis of these thermodynamic quantities, as they capture the essential behavior of each. Additionally, parameterizing in terms of temperature, mass, and chemical potential facilitates the computation of the required integrals, providing a useful framework for applying approximations, such as $m \ll T$, which corresponds to the ultra-relativistic case.\linebreak Generally, the integrals (\ref{mathcal NC})--(\ref{mathcal PC}) cannot be expressed in terms of standard \mbox{special functions} and require numerical methods for evaluation under certain limiting conditions.\linebreak However, in some cases discussed later, these integrals can be solved analytically.

\subsubsection{Non-Relativistic Case}\label{subsec:Non-relativistic case}

In the specific case of approximated Maxwell--Boltzmann statistics, the integrals (\ref{mathcal NC})--(\ref{mathcal PC}) can be computed and expressed in terms of special functions.  The corresponding integrals can be formulated as
\begin{align}
\mathcal{N}_{i}^{\textrm{MB}}\left(\frac{T_i}{T},~\frac{m_i}{T_i},~\frac{\mu_i}{T_i}\right)&=\frac{\pi^4}{45}\frac{T}{T_i}\mathcal{P}^{\textrm{MB}}_{i}\left(\frac{T_i}{T},~\frac{m_i}{T_i},~\frac{\mu_i}{T_i}\right)=g_i\,\left(\frac{T_i}{T}\right)^3\,\left(\frac{m_i}{T_i}\right)^2e^{\mu_i/T_i}K_2(z_i)\ ,\\
\mathcal{E}^{\textrm{MB}}_{i}\left(\frac{T_i}{T},~\frac{m_i}{T_i},~\frac{\mu_i}{T_i}\right)&=\frac{45}{\pi^4}\,\left(\frac{T_i}{T}\right) \left(1 + \frac{K_1(z_i)}{3K_2(z_i)}\right) \mathcal{N}_{i}^{\textrm{MB}}\, .
\end{align}


In this context, $K_n(z)$ denotes a modified Bessel function of the second kind, and the superscript signifies the reinterpretation of the integral functions (\ref{mathcal NC})--(\ref{mathcal PC}) in terms of the Maxwell--Boltzmann (MB) distribution function. This reinterpretation involves the omission of the $\pm 1$ term present in the original functions. As previously highlighted, in the absence of degenerate fermions and a Bose condensate, the approximation introduced by this statistical approach only results in a minor quantitative deviation compared to the exact Fermi--Dirac (FD) and Bose--Einstein (BE) statistics. Consequently, in the pertinent physical limit of interest, where the particle species is non-relativistic ($m_i\gg T_i$) and $m_i\gg T_i+\mu_i$, this approximation becomes exact. The second condition gives rise to occupation numbers significantly below unity, indicating a dilute system, a circumstance that aligns with the assumption made earlier. This condition is commonly met in cosmological contexts when the first one is satisfied. However, it is notably violated in systems characterized by high density, such as white dwarfs and neutron stars. Consequently, $\mathcal{N}_i^j$, $\mathcal{E}_i^j$, and $\mathcal{P}_i^j$ become independent of the superscript, denoted as $j=\textrm{MB},~+,~-$. Taking into consideration that for $z\gg 1$ ($m\gg T$), $K_2(z_i)=\sqrt{\pi/(2z_i)}e^{-z}+O(1/z_i)$ and $\left(1 + \frac{K_1(z_i)}{3K_2(z_i)}\right) = \frac{1}{2} + \frac{z_i}{3}$,\linebreak a straightforward derivation follows
\begin{eqnarray}
\mathcal{N}_{i}^{\pm}\left(\frac{T_i}{T},~\frac{m_i}{T_i},~\frac{\mu_i}{T_i}\right) 
&=&g_i\,\sqrt{\frac{\pi}{2}}\,\left(\frac{T_i}{T}\right)^3\,\left(\frac{m_i}{T_i}\right)^\frac{3}{2}e^{-\frac{m_i-\mu_i}{T_i}}, \\ 
\mathcal{P}_{i}^{\pm}\left(\frac{T_i}{T},~\frac{m_i}{T_i},~\frac{\mu_i}{T_i}\right)
&=&\frac{45}{\pi^4}\,\left(\frac{T_i}{T}\right)\mathcal{N}_{i}^{\pm}\left(\frac{T_i}{T},~\frac{m_i}{T_i},~\frac{\mu_i}{T_i}\right),\\ 
\mathcal{E}_{i}^{\pm}\left(\frac{T_i}{T},~\frac{m_i}{T_i},~\frac{\mu_i}{T_i}\right) &=& 
\frac{45}{\pi^4}\,\left(\frac{T_i}{T}\right)\,\left[\frac{1}{2}+\left(\frac{m_i}{3T_i}\right)\right]\mathcal{N}_{i}^{\pm}\left(\frac{T_i}{T},~\frac{m_i}{T_i},~\frac{\mu_i}{T_i}\right), 
\end{eqnarray}
leading to the non-relativistic number density ($n_i$), energy density ($\rho_i$), and pressure ($p_i$)
\begin{align}
n_i&=g_i\,\left(\frac{m_iT_i}{2\pi}\right)^{3/2}e^{-\frac{m_i-\mu_i}{T_i}} \, ,\label{non-relativistic number density}\\
\rho_i&=m_in_i+\frac{3}{2}n_iT_i \label{non-relativistic energy density} \, ,\\
p_i&=n_iT_i\ll \rho_i \label{non-relativistic pressure}
\,,
\end{align}
respectively.

In the scenario of massless particles ($m_i=0$), it is possible to express the integrals in Equations (\ref{mathcal NC})--(\ref{mathcal PC}) in terms of special functions. The results are as follows:
\begin{align}
\mathcal{N}^{\pm}_{i}\left(\frac{T_i}{T},~0,~\frac{\mu_i}{T_i}\right)&=\mp 2g_i\,\left(\frac{T_i}{T}\right)^3Li_3(\mp e^{\mu_i/T_i})\label{polylogarithmic mathcal N}\\
\mathcal{E}^{\pm}_i\left(\frac{T_i}{T},~0,~\frac{\mu_i}{T_i}\right)&=\mathcal{P}_i^{\pm}\left(T,~0,~\frac{\mu_i}{T_i}\right)=\mp\frac{90}{\pi^4}g_i\,\left(\frac{T_i}{T}\right)^4Li_4(\mp e^{\mu_i/T_i})\label{polylogarithmic mathcal E and P}\,,
\end{align}
where $Li_s(x)$ are polylogarithmic functions. For small but not vanishing chemical potentials, these can be expressed as
\begin{adjustwidth}{-\extralength}{0cm}
\begin{align}
\mathcal{N}_i^+\left(\frac{T_i}{T},~0,~\frac{\mu_i}{T_i}\right)&=g_i\,\left(\frac{T_i}{T}\right)^3\left[\frac{3\zeta(3)}{2}+\frac{\pi^2}{6}\,\left(\frac{\mu_i}{T_i}\right)+\log(2)\,\left(\frac{\mu_i}{T_i}\right)^2+...\right]\,,
\\
\mathcal{N}_i^-\left(\frac{T_i}{T},~0,~\frac{\mu_i}{T_i}\right)&=g_i\,\left(\frac{T_i}{T}\right)^3\left[2\zeta(3)+\frac{\pi^2}{3}\,\left(\frac{\mu_i}{T_i}\right)+\frac{1}{2}\,\left(3-2\log\left(\frac{\mu_i}{T_i}\right)\right)\,\left(\frac{\mu_i}{T_i}\right)^2+...\right]\,,
\end{align}
\end{adjustwidth}

\begin{align}
\mathcal{E}_i^+\left(\frac{T_i}{T},~0,~\frac{\mu_i}{T_i}\right)&=\mathcal{P}_i^+\left(\frac{T_i}{T},~0,~\frac{\mu_i}{T_i}\right)
	\nonumber \\
&=g_i\,\left(\frac{T_i}{T}\right)^4 \left[\frac{7}{8}+\frac{135\zeta(3)}{2\pi^4}\,\left(\frac{\mu_i}{T_i}\right)+\frac{15}{4\pi^2}\,\left(\frac{\mu_i}{T_i}\right)^2+...\right]\,,
\\
\mathcal{E}_i^-\left(\frac{T_i}{T},~0,~\frac{\mu_i}{T_i}\right)&=\mathcal{P}_i^-\left(\frac{T_i}{T},~0,~\frac{\mu_i}{T_i}\right)
	\nonumber \\
&=g_i\,\left(\frac{T_i}{T}\right)^4\left[1+\frac{90\zeta(3)}{\pi^4}\,\left(\frac{\mu_i}{T_i}\right)+\frac{15}{2\pi^2}\,\left(\frac{\mu_i}{T_i}\right)^2+...\right]
\,.
\end{align}

\subsubsection{Relativistic Case}\label{subsec:Relativistic case}

In the pertinent physical limit of interest, where the species are relativistic, $T_i\gg m_i$, one can consider vanishing chemical potentials ($\mu_i=0$), which is a reasonable approximation given that $|\mu| \ll T$, and a straightforward derivation can be made from Equations \eqref{polylogarithmic mathcal N} and \eqref{polylogarithmic mathcal E and P} by setting $\mu_i=0$ or from Equations \eqref{mathcal NC} and \eqref{mathcal EC} using $m_i/T_i \to 0 $ and $\mu_i/T_i \to 0$. With this being said, the relativistic case presents  
\begin{equation}
\mathcal{N}_i^{\pm}\left(\frac{T_i}{T},0,~0\right)=2\zeta(3)g^n_{\textrm{eff}_i}\,\left(\frac{T_i}{T}\right)^3, 
\end{equation}
\begin{equation}
\mathcal{E}_i^{\pm}\left(\frac{T_i}{T},0,~0\right)=\mathcal{P}\left(\frac{T_i}{T},0,~0\right)=g_{\textrm{eff}_i}\left(\frac{T_i}{T}\right)^4 \, ,
\end{equation}
that leads to the relativistic number density of species $i$ and energy density of relativistic particle species $i$
\begin{equation}
    n_i=\frac{\zeta(3)}{\pi^2}g^n_{\textrm{eff}_i}T_i^3 \, ,
\end{equation}
\begin{equation}
\rho_i=\frac{\pi^2}{30}g_{\textrm{eff}_i}T_i^4=3p_i\,,\label{rho_relativistic}
\end{equation}
where $g^n_{\textrm{eff}_i}=3/4g_i$ and $g_{\textrm{eff}_i}=7g_i/8$ if the particle species under consideration is a fermion and $g^n_{\textrm{eff}_i}=g_{\textrm{eff}_i}=g_i$ if the particle species is a boson. 

Based on the results for both non-relativistic and relativistic scenarios, the total energy density can be derived through two key considerations. First, by examining equilibrium conditions, the total energy density can be approximated by focusing mainly on relativistic species, as the contributions from non-relativistic species are exponentially smaller~\cite{Kolb:1990vq}. Next, we assume that the medium has a temperature $T$ equal to that of the photon gas. With these assumptions, the total energy density can be calculated using Equation \eqref{total energy density}, resulting in~\cite{Kolb:1990vq}
\begin{equation}\label{total rho_relativistic}
    \rho = \frac{\pi^2}{30} g_\ast T^4 \, ,
\end{equation}
where
\begin{equation}
g_{\ast}(T)=\sum_{i=\textrm{bosons}}g_i\,\left(\frac{T_i}{T}\right)^4+\frac{7}{8}\sum_{i=\textrm{fermions}}g_i\,\left(\frac{T_i}{T}\right)^4\,,
\end{equation}
counts the effectively massless degrees of freedom~\cite{Kolb:1990vq} (i.e., species with mass $m_i \ll T$).\linebreak The sum over particle species can be divided into two types of contributions. The first type, denoted as $g_\ast^{equilibrium}$, accounts for relativistic species that are in thermal equilibrium with the medium, meaning their temperature is $T_i = T$. This case provides
\begin{equation}
g_\ast^{equilibrium}=\sum_{i=\textrm{bosons}}g_i+\frac{7}{8}\sum_{i=\textrm{fermions}}g_i \, .
\end{equation}

The second contribution, $g_\ast^{decoupled}$, occurs when the species are not in thermal equilibrium with the photon gas, $T_i \neq T$, 
\begin{equation}
g_\ast^{decoupled}(T)=\sum_{i=\textrm{bosons}}g_i\,\left(\frac{T_i}{T}\right)^4+\frac{7}{8}\sum_{i=\textrm{fermions}}g_i\,\left(\frac{T_i}{T}\right)^4\, .
\end{equation}

Hence, these outcomes hold true for any given ultra-relativistic particle species $i$ exhibiting a small chemical potential ($\mu_i\ll T_i$) and possessing a distribution function with an equilibrium shape. Notably, in the case of the number density, the obtained results align with Equation~\eqref{number density of decoupled massless particle species 1a} as expected.

\subsubsection{Net Particle Number}\label{subsec:Net particle number}

The comparison between a species and its antiparticle is often of interest, and in a state of complete equilibrium, its excess can be easily computed. With chemical equilibrium implying that the chemical potential of a particle species $i=i^{+}$ is equal in magnitude and symmetric in sign to the chemical potential of its antiparticle $i=i^{-}$, denoted as $\mu\equiv \mu_{i^+}=-\mu_{i^-}$, the net density number of a particle species $i^{+}$ over its antiparticle $i^{-}$ can be computed by using Equation~\eqref{number density} giving~\cite{Kolb:1990vq}
\begin{adjustwidth}{-\extralength}{0cm}
\begin{eqnarray}
n_{i^+}-n_{i^-}=\left\{
\begin{array}{ll}\frac{g_iT_i^3}{6\pi^2}\,\left[\pi^2\,\left(\frac{\mu}{T_i}\right)+\left(\frac{\mu}{T_i}\right)^{3}\right]\,,&\textrm{for relativistic fermions}\,(T_i\gg m_i)
\label{excess of particles over antiparticles 1 a} \\
\frac{g_iT_i^3}{6\pi^2}\,\left[2\pi^2\left(\frac{\mu}{T_i}\right)-\frac{1}{3}\,\left(\frac{\mu}{T_i}\right)^{3}\right]\,,&\textrm{for relativistic bosons}\,(T_i\gg m_i)\\
2g_i\,\left(\frac{m_iT_i}{2\pi}\right)^{3/2}\sinh\left(\frac{\mu}{T_i}\right)e^{\left(-\frac{m_i}{T_i}\right) }\,,&\textrm{for all non-relativistic species}\,(T_i\ll m_i) \label{excess of particles over antiparticles 1 b}  \,
\end{array}
\right.
\end{eqnarray}
\end{adjustwidth}

It is also often useful to express these quantities in terms of small chemical potentials, in which case for $\vert\mu\vert<m$ (no Bose condensation)
\begin{eqnarray}
n_{i^+}-n_{i^-}=\frac{g_i a^{\pm c}}{6}T_i^3\,\left(\frac{\mu_i}{T_i}\right)\alpha^{\pm c}_i\left(\frac{m_i}{T_i}\right)\,,\label{excess of particles over antiparticles 1 b for small chemical potentials}
\end{eqnarray}
where
\begin{eqnarray}
\alpha^{\pm c}_i\left(z_i=\frac{m_i}{T_i}\right)\equiv\frac{6}{\pi^2a^{\pm c}} \int_{z_i}^{\infty}u\sqrt{u^2-z_i^2}\frac{e^u}{(e^u\pm c)^2}\,\mathrm{d}u\,,\qquad(u=E_i/T)\,, \label{function alpha 1}
\end{eqnarray}
with $a^{+1}=1$, $a^{-1}=a^{\pm 0}=2$ and the superscripts $+1$, $-1$ and $\pm 0$ standing respectively for FD, BE, and MB statistics. Note that in the massless limit ($m_i=0$), $\alpha^{\pm c}(0)=1$.

\subsection{Entropy in the Expanding Universe}

Entropy holds a crucial position in the thermodynamic description of the universe,\linebreak particularly due to its relationship with the scale factor. Within the field of thermodynamics, entropy is also fundamental to the discipline's core principles. This is especially true in scenarios involving a varying number of particles, where entropy's significance is highlighted by the general formulation given in Equation \eqref{Entropia}. This formulation is structured around the understanding that energy and the number of particles are extensive properties, scaling proportionally with the volume of the system, whereas temperature and pressure manifest as local characteristics independent of volume. Consequently, entropy itself is classified as an extensive property. It is beneficial, therefore, to reformulate the fundamental thermodynamic equation by expressing the energy, number of particle, and entropy in terms of their densities within a cosmological volume $V$. This reformulation leads to the\linebreak  following expression:
\begin{equation}
\vspace{-3pt}
(Ts-\rho-p+\mu n)dV+(Tds-d\rho+\mu dn)V=0\,,
\end{equation}
where $\rho=E/V$, $n=N/V$ and $s=S/V$. This relation is valid both for the entire system and any of its parts and so, using it for a region of constant volume inside the system, it is possible to obtain $Tds=d\rho-\mu dn$, and using it subsequently for the entire system, gives the entropy density~\cite{Rubakov:2017xzr} 
\begin{equation}\label{entropy density geral}
s=\sum_i\frac{\rho_i+p_i-\sum_i\mu_i n_i}{T_i}=\frac{2\pi^2}{45}\mathcal{S}_{\ast}\left(\left\{\frac{T_i}{T}\right\},~\left\{\frac{m_i}{T_i}\right\},~\left\{\frac{\mu_i}{T_i}\right\}\right)\,T^3\,,
\end{equation}
where $S_{\ast}(\{T_i/T\},\, \{m_i/T\},\,\{\mu_i/T\})=\sum_iS_i^{\pm}(T_i/T,\, m_i/T,\,\mu_i/T)$ with $S^{\pm}_i$ defined as
\begin{equation}
S_i^{\pm}\left(\frac{T_i}{T},~\frac{m_i}{T_i},~\frac{\mu_i}{T_i}\right)\equiv \frac{T}{T_i}\left(\frac{3}{4}\mathcal{E}_i^{\pm}+\frac{1}{4}\mathcal{P}_i^{\pm}\right)-\frac{45}{4\pi^4}\frac{\mu_i}{T_i}\mathcal{N}_i^{\pm} \, ,
\end{equation}
represents the contributions of each species $i$ to the effective degrees of freedom within entropy. Equation~\eqref{entropy density geral} corresponds to the general expression without assuming any specific case, where the computation of such a general case requires numerical methods. However, the relativist and non-relativist cases can be calculated analytically. 

In the relativistic case, considering $T_i\gg m_i$ and vanishing chemical potentials, the expression for $\mathcal{S}_i^{\pm}$ is given by
\begin{equation}
\mathcal{S}_i^{\pm}\left(\frac{T_i}{T},~0,~0\right)=\frac{g_{\textrm{eff}_i}}{2\zeta(3)g_{\textrm{eff}_i}^n}\mathcal{N}_i^{\pm}\left(\frac{T_i}{T},~0,~0\right)
=g_{\textrm{eff}_i}\left(\frac{T_i}{T}\right)^3 \, ,
\label{entropy density in relativistic particle species}
\end{equation}
which in turn allows to compute the total energy density for relativistic species 
\begin{equation}
    s_{relativistic} = \sum_i \frac{2\pi^2}{45}g_{\textrm{eff}_i}T_i^3 \, .
\end{equation}

Another way to obtain this result is to use directly the first equality of Equation~\eqref{entropy density geral} with $ \mu = 0 $ and considering that in the relativistic scenario, the energy density and pressure have the relation $\rho = 3p$, allowing in combination with Equation~\eqref{rho_relativistic} to write the entropy density for relativistic species as
\begin{equation}
    s_{relativistic} = \sum_i \frac{\rho_i + p_i}{T_i} = \frac{4}{3} \sum_i \frac{\rho_i}{T_i} = \sum_i \frac{2\pi^2}{45}g_{\textrm{eff}_i}T_i^3 \, .
\end{equation}

For the non-relativistic case, one considers $m_i\gg T_i$ satisfying the condition\linebreak $m_i-\mu_i \gg T$ that allows to compute $\mathcal{S}_i^{\pm}$ as
\begin{equation}
\mathcal{S}_i^{\pm}\left(\frac{T_i}{T},~\frac{m_i}{T_i},~\frac{\mu_i}{T_i}\right)=\frac{45g_i}{4\pi^4}\sqrt{\frac{\pi}{2}}\left(\frac{5}{2}+\frac{m_i-\mu_i}{T_i}\right)\,
\left(\frac{T_i}{T}\right)^3\,
\left(\frac{m_i}{T_i}\right)^\frac{3}{2}e^{\frac{\mu_i-m_i}{T_i}},
\end{equation}
leading to the following total entropy density for non-relativistic species 
\begin{equation}
    s_{non-relativistic} = \sum_i n_i\left[ \frac{5}{2} + \ln\left( \frac{g_i}{n_i} (\frac{m_i T_i}{2\pi})^{\frac{3}{2}}\right)\right] \, ,
\end{equation}
where Equation~\eqref{non-relativistic number density} is used to express the entropy density in terms of the  number density.

\textls[-15]{Once more, this result can be obtained directly by using the first equality of \mbox{Equation~\eqref{entropy density geral},}} Equation~\eqref{non-relativistic energy density}, and Equation~\eqref{non-relativistic pressure}, yielding~\cite{Rubakov:2017xzr}
\begin{equation}
    s_{non-relativistic} = \sum_i \frac{5}{2} n_i + \frac{m_i -\mu_i}{T} n_i \, .
\end{equation}

To calculate the total entropy density, one can approximate that relativistic species will dominate the contribution when comparing the results for relativistic and non-relativistic cases. Therefore, using Equation \eqref{rho_relativistic}, the total entropy density can be expressed as~\cite{Kolb:1990vq}
\begin{equation}
    s = \sum_i \frac{\rho_i + p_i}{T_i} = \frac{2\pi^2}{45} g_{\ast S} T^3 \, ,
\end{equation}
where 
\begin{equation}
    g_{\ast S}  =\sum_{i=\textrm{bosons}}g_i\,\left(\frac{T_i}{T}\right)^3+\frac{7}{8}\sum_{i=\textrm{fermions}}g_i\,\left(\frac{T_i}{T}\right)^3\, . 
\end{equation}

A notable point to consider is that when all relativistic species are in thermal equilibrium—meaning they share the same temperature, it is a good approximation to assume that $g_{\ast S }$ equals $g_\ast$~\cite{Kolb:1990vq}. The entropy of all relativistic species remains conserved as long as their distribution functions stay thermal. In this scenario, any entropy production from non-equilibrium processes is negligible compared to the total entropy, which is overwhelmingly dominated by the relativistic species. Therefore, treating the expansion of the universe as adiabatic is an excellent approximation. In this context, entropy serves as a reliable tool for tracking the evolution of the scale factor in relation to temperature, which is then given by the following expression:
\begin{equation}
a\propto g_{\ast S}^{-1/3}(T)\,T^{-1}\,.
\end{equation}

It is worth mentioning that when a particle species decouples, the temperature of the remaining equilibrium radiation decreases at a slower rate. This occurs because the number of degrees of freedom within the equilibrium radiation decreases~\cite{Kolb:1990vq}. Additionally, the temperature of the decoupled species decreases as $T^{-1}$, in contrast to the $g_{\ast}^{-1/3}(T)T^{-1}$ dependence observed in equilibrium radiation. This discrepancy arises because the entropy density of the decoupled species remains conserved independently, and therefore, it does not interact with or contribute to the degrees of freedom of the equilibrium radiation.
 
These results clearly demonstrate the significant impact of the second law of thermodynamics on our understanding of the primordial universe. According to this law, the entropy of any closed system must increase, remaining constant only during equilibrium or adiabatic processes, where the system evolves gradually while maintaining thermal equilibrium. In the context of an expanding universe, applying the fundamental principles of thermodynamics to a comoving volume, defined as $V\equiv a^3$, reveals that during thermal equilibrium, the following condition holds:
\begin{eqnarray}
\frac{d(a^3s)}{dt} &=& \frac{d}{dt}\,\left[\frac{a^3}{T}(\rho+p-\mu n)\right]
	\nonumber \\
&=&\frac{a^3}{T}\,\left[3H(\rho+p)+\frac{d\rho}{dt}+\frac{dp}{dt}-s\frac{dT}{dt}-n\frac{d \mu}{dt}-\mu\left(\frac{dn}{dt}-3H n\right)\right]\,,
\end{eqnarray}
where the subscript $i$ labeling the particle species and the corresponding summation is omitted, and recognizing that for vanishing chemical potentials $\dot{p}=s\dot{T}$, it follows from the conservation of energy Equation \eqref{Conservation} that for vanishing chemical potentials,
\begin{equation}
\frac{d(a^3s)}{dt}
=0\label{conservaçao da entrpoia }\,.
\end{equation}

It becomes clear that the equation governing energy conservation can be interpreted directly in terms of entropy. When equilibrium is maintained and the chemical potentials for particle species vanish, this equation transforms into one that describes the conservation of entropy within a comoving volume, expressed as $S = a^3s$. This insight has broader implications.
First, it applies to any particle species with a distribution function of the form $f = f^{\textrm{eq}}(E/T)$, where $T(t)$ is a time-dependent temperature, as long as they follow the energy conservation equation. This generalization also holds true for any decoupled stable particle species that maintain a thermal distribution with their own distinct temperatures.
Second, by revisiting the fundamental thermodynamic relation, we can introduce chemical potentials not only for individual particle species but also for conserved quantum numbers. When $dN_i$ represents the differentials of such conserved quantum numbers, $dN_i = 0$ whenever these quantum numbers are preserved. Thus, even when chemical potentials are non-zero, the previously derived results remain valid.

From this discussion, two important conclusions can be drawn. First, for\linebreak a universe undergoing adiabatic expansion, or more precisely, when entropy within a comoving volume is conserved, a direct relationship can be established between the universe's expansion (or equivalently, its redshift) and its cooling, characterized by the\linebreak following parameters:
\phantom{0}\vspace{-12pt}
\begin{align}
\vspace{-3pt}
a\left(T,\left\{\frac{T_i}{T}\right\},\left\{\frac{m_i}{T_i}\right\},\left\{\frac{\mu_i}{T_i}\right\}\right)&=cte\, \left(\frac{45}{2\pi^2}\right)^{1/3}\mathcal{S}_{\ast}^{-1/3}\left(\left\{\frac{T_i}{T}\right\},\left\{\frac{m_i}{T_i}\right\},\left\{\frac{\mu_i}{T_i}\right\}\right)\,T^{-1}\, ,\label{entropy conservation scale factor temperature relation}\\
(1+z_1)\,T_0&=\left(\frac{\mathcal{S}_{\ast}^{1/3}(\{T_i/T_1\},\{m_i/T_i\},\{\mu_i/T_i\})}{\mathcal{S}_{\ast}^{1/3}(\{T_i/T_0\},\{m_i/T_i\},\{\mu_i/T_i\})}\right)\,T_1\,,\label{redshift  and cooling characterization}
\end{align}
and the number of a given species $i$ in a comoving volume, $V_i=a^3n_i$, is proportional to the number density of the species $N$ divided by $s$, that is, $V_i=cte.\times n_i/s= cte\times N$, where $N$ is conveniently defined as
\begin{equation}
N\equiv \frac{n_i}{s}\,.\label{number of species per entropy density}
\end{equation} 

Subsequently, from Equation \eqref{entropy conservation scale factor temperature relation} it is possible to establish the differential relation:
\begin{align}
\frac{da}{a}&=H\mathrm{d}t=-\left(\frac{\mathrm{d}\mathcal{S}_{\ast}}{3\mathcal{S}_{\ast}}+\frac{dT}{T}\right)\quad \Rightarrow \quad H=-\left(\frac{\dot{\mathcal{S}_{\ast}}}{3\mathcal{S}_{\ast}}-\frac{\dot{T}}{T}\right)\,,
\end{align}
which in turn, combined with the dynamical Equation~\eqref{Friedmann} for the flat model, written in terms of $\rho=\pi^2\mathcal{E}_{\ast}T^4/30$, can be used to find the \emph{time--temperature relation} 
\begin{equation}
t=-\int \left(\frac{8\pi^3 G_N}{90}\mathcal{E}_{\ast}T^4\right)^{-1/2}\left(\frac{\mathrm{d}\mathcal{S}_{\ast}}{3\mathcal{S_{\ast}}}+\frac{dT}{T}\right)\,.\label{exact time-temperature relation}
\end{equation}
which relates the age of the universe with the temperature.

\subsection{Baryons in the Universe}

Building on the conclusions drawn in the preceding section, it is evident that the conservation of entropy per comoving volume can be employed to introduce quantitative, time-independent characteristics of asymmetries in conserved quantum \mbox{numbers~\cite{A.D. Dolgov,Rubakov:2017xzr}}. Specifically, this framework allows for the examination of baryon asymmetry in the universe, characterized by the presence of baryonic matter and the apparent absence of antimatter for practical purposes. In the absence of baryon number-violating processes, the baryon asymmetry must be conserved within a comoving volume, thereby ensuring that

\begin{equation}
(n_b-n_{\bar{b}})a^3=cte\,,
\end{equation}
where $n_{b}$ and $n_{\bar{b}}$ are the number densities of baryons and antibaryons, respectively, and  so, from Equation~(\ref{number of species per entropy density}), it can be immediately seen that the ratio
\begin{equation}
B\equiv \frac{n_B}{s}\,,
\end{equation}
where $ n_B \equiv n_b - n_{\bar{b}}$ represents a time-independent characteristic of baryon asymmetry, assuming the expansion of the universe remains isotropic. This condition is typically satisfied with high precision throughout most of the universe's history. In the post-photon decoupling, the number of non-interacting photons in a comoving volume remains constant, with $ T_{\gamma} \propto a^{-1}(t) $ and $ n_{\gamma} \sim a^{-3}(t) $. Consequently, after this point, as long as there is no further violation of the baryon number, the ratio $ n_{B}/n_{\gamma} $ remains unchanged.

At higher temperatures, however, the annihilation of massive particles occurs when the temperature drops below their respective masses. These annihilations heat the primordial plasma, increasing the photon number density. Therefore, it is more practical to introduce the quantity $ B $, which remains effectively constant in a state of thermal equilibrium throughout the expansion. Within the SM of particle physics, in the absence of new long-lived particles, there is no further transfer of entropy to the photons after the $ e^+e^- $ annihilation and as a result, $ n_{\gamma}a^3 $ also remains constant. Therefore, when lower temperatures are reached, specifically around $ T \lesssim 1 $ MeV, the \emph{baryon-to-photon} ratio is used
\begin{equation}
\eta\equiv \frac{n_B}{n_{\gamma}}\,,
\end{equation}
which, for this temperature range, also remains constant and can easily be related to the quantity $B$ by
\begin{equation}
\eta=\frac{s}{n_{\gamma}}B=\frac{\left(\frac{2\pi^2}{45}\mathcal{S}_{\ast, 0}\right)}{\left(\frac{2\zeta(3)}{\pi^2}\right)B} \approx \frac{B}{0.14}\,.
\end{equation}

The \emph{baryon-to-photon} ratio is roughly equal to $10 B$ so that the two quantities practically coincide with each other. At higher temperatures, the baryon-to-photon ratio may differ by one to two orders of magnitude due to the contribution of heavier particles to the entropy density, resulting in the dilution of the baryon-to-photon ratio by the same amount. Additional sources of dilution include possible first- and second-order phase transitions in the early universe and the out-of-equilibrium decay of unstable particles. These processes could significantly diminish the original baryon asymmetry characterized by $n_B/s$. All evidence suggests that the present-day baryon asymmetry $n_B/s \sim 10^{-11}$\mbox{~\cite{Planck:2018vyg,Fields:2019pfx,ParticleDataGroup:2018ovx,ParticleDataGroup:2020ssz,ParticleDataGroup:2022pth,BB,WMAP}}. This meager value has profound implications for the thermal history of the universe. Most importantly, this result supports the approximation made to compute the total entropy density, as it allows us to conclude that the number density of non-relativistic species is always small in the cosmic medium at low temperatures when protons and neutrons \mbox{are non-relativistic}.

From a phenomenological perspective, the observed baryonic asymmetry presents a significant gap in our understanding of the primordial universe. According to the Standard Model (SM) of particle physics, our best current theory, the Big Bang should have produced equal amounts of particles and antiparticles. This would have led to their complete annihilation, leaving behind a universe filled only with photons~\cite{Riotto:1999yt}.\linebreak However, this prediction contrasts sharply with what we actually observe. The existence of baryonic asymmetry suggests that a physical mechanism must have created this imbalance at some point during the universe's evolution. The processes responsible for generating this asymmetry are known as baryogenesis mechanisms and offer the best explanation for this issue.

The foundational work in baryogenesis was published in 1967 when A.D. Sakharov proposed that the baryon asymmetry might not result from unnatural initial conditions but could instead be explained through microphysical laws~\cite{Sakharov}. These laws suggest that an initially symmetric universe could dynamically develop the observed asymmetry, providing a theoretical framework for understanding this phenomenon within particle physics and cosmology.
Sakharov outlined a ``recipe'' for generating this asymmetry, which has become fundamental to the study of baryonic asymmetry. The three conditions defined by Sakharov are as follows: 

\begin{enumerate}
\item Baryon number violation;
\textls[-15]{\item Violation of C (charge conjugation symmetry) and CP (the composition of parity and C);}
\item Departure from the equilibrium.
\end{enumerate}  

Each of these conditions plays a crucial role in successfully generating asymmetry. The first condition allows for the creation of baryons by ensuring that the baryon number is not conserved. The second condition favors the production of matter over antimatter. The third condition helps maintain the asymmetry by suppressing any reverse processes when the system moves out of equilibrium.
However, it was later discovered that while the Sakharov conditions are effective in creating asymmetry, they are not strictly necessary. It is possible to develop successful baryogenesis mechanisms that do not meet all three conditions~\cite{Cohen:1987vi,Lambiase:2013haa,Dolgov:1991fr,Dolgov:1997qr,Arbuzova:2016cem}.

The majority of baryogenesis mechanisms rely on physics beyond the SM (for more details, see~\cite{Dine:2003ax,Cline:2006ts,Bodeker:2020ghk,Cui:2015eba,Allahverdi:2012ju}), showing how this asymmetry can be a portal to new physics. The need for physics beyond the SM arises not only as a potential solution to the asymmetry problem, where Grand Unified Theories (GUTs) were initially considered due to their natural incorporation of baryon number violation, but also due to the limited CP violation of the SM and the shortcomings of electroweak baryogenesis, a mechanism that relies solely on the SM~\cite{Trodden:1998ym,Trodden:1998qg,Morrissey:2012db,Cline:2017jvp,Pereira:2023xiw}.


Since electrons and nucleons are present today, this implies that during the primordial era, the chemical potential for baryons, $\mu_B$, was non-zero and positive. Moreover, because the universe is electrically neutral and electrons are the lightest negatively charged particles, the excess of electrons over positrons must balance with the excess of nucleons over antinucleons. For each particle species, $\mu(T)$ can be determined based on the conserved quantities and the current number of fermions (electrons and nucleons). This analysis shows that in the early universe, $\mu \ll T$ when $T \gg m$. Specifically, the baryon chemical potential is very small at temperatures around $T \gtrsim 200$ MeV~\cite{Rubakov:2017xzr}. At lower temperatures, the net baryon density is so negligible compared to the photon density that this chemical potential can be effectively disregarded in calculations of total thermodynamic quantities~\cite{ParticleDataGroup:2018ovx,ParticleDataGroup:2020ssz,ParticleDataGroup:2022pth}.\linebreak The same reasoning applies to other chemical potentials (noting that, for practical purposes, photons always have zero chemical potential).

However, it is important to emphasize that while at temperatures $200 \text{ MeV} \lesssim T \lesssim 100 \text{ GeV}$, baryon number $B$, lepton numbers $L_e$, $L_\mu$, $L_\tau$, and electric charge $Q$ are conserved quantum numbers, at higher temperatures, this conservation may not hold true~\cite{Rubakov:2017xzr}.\linebreak In most scenarios above $T \sim 100 \text{ GeV}$, there is no violation of the baryon number, although some models propose such violations even at lower temperatures~\cite{Rubakov:2017xzr}. Generally, it is assumed that the universe was initially symmetric in matter, and the observed\linebreak asymmetry was generated by baryon number-violating processes occurring out of equilibrium. In this situation, the chemical potential for baryons would initially vanish, with the\linebreak aforementioned processes producing the small asymmetry observed today. Consequently, the chemical potential for baryons, and therefore for electrons, could typically be disregarded. Nevertheless, this view can be contested. For example, if the baryon asymmetry in the early universe was higher than it is today and was subsequently diluted to its present value by processes such as entropy generation or baryon-violating processes occurring in equilibrium, the previous conclusion would not hold.

\subsection{Equilibrium in the Expanding Universe\label{subsec:Equilibrium in the Expanding Universe}}

Given the expanding nature of the universe, maintaining equilibrium under such conditions is complex and not straightforward. Consequently, a variety of cases and situations must be considered, analyzed, and discussed.
The previous results suggest that in the usual cosmological framework, especially at the high temperatures typical of the early universe, chemical potentials are expected to be very small and can be neglected to a good approximation. This simplification greatly simplifies calculations, as thermal distribution functions and all thermodynamic quantities then depend solely on temperature. By ignoring chemical potentials, we can compare the asymptotic forms of $\mathcal{N}_i^{\pm}$, $\mathcal{E}_i^{\pm}$, and $\mathcal{P}_i^{\pm}$ in both relativistic and non-relativistic limits. Specifically, the term $z_i^{3/2}\exp(-z_i)$, common to $\mathcal{N}_i^{\pm}$, $\mathcal{E}_i^{\pm}$, and $\mathcal{P}_i^{\pm}$ in the non-relativistic limit, reaches a maximum at $z_i = 1.5$, approximately 0.401, and becomes exponentially suppressed for higher $z_i$ values.

In equilibrium, the contributions from non-relativistic particles to these quantities are exponentially smaller compared to those from relativistic particles. This difference arises because, in the non-relativistic limit, particle densities, energies, and pressures decrease exponentially with temperature, whereas in the relativistic limit, they decrease more gradually. Thus, when equilibrium is maintained among the primary components of the universe's matter content, it is usually a good approximation to include only relativistic particles in the total thermodynamic quantities. Even the contributions from semi-relativistic particles can often be disregarded, as they introduce only a minor error, which is generally insignificant for most situations. Using this approximation, the Friedmann equation can be expressed as given by Equation~\eqref{total rho_relativistic}:
\begin{equation}
    H^2 = \frac{8 \pi^3 G_N}{90} g_\ast(T) T^4  
\end{equation}
leading to
\begin{equation}\label{radiation thermal expansion rate}
H=1.66g_{\ast}^{1/2}(T)\frac{T^2}{m_{Pl}}\,,
\end{equation}
where $m_{Pl}=G_N^{-1/2}=1.22\times 10^{19}$ GeV is the Planck mass. It should be noted that $ g_{\ast}(T) $ is inherently temperature dependent because as temperature varies, particle species may transition between relativistic and non-relativistic states, thereby contributing to or being excluded from the sum respectively. The specific value of $ g_{\ast}(T) $ at a given temperature is also influenced by the underlying particle physics model. Generally, considering its temperature dependence, $ \sqrt{g_{\ast}} $ typically ranges between 2 and 20. For example, within the standard $\textrm{SU}(3)\times \textrm{SU}(2)\times \textrm{U}(1)$ model, $ g_{\ast}(T) $ can be specified for temperatures up to $\textrm{O}(100)$ GeV. However, at higher temperatures, $ g_{\ast}(T) $ becomes model dependent. In the minimal $\textrm{SU}(5)$ model for $ T \gtrsim 10^{15} $ GeV, $ g_{\ast}(T) = 160.75 $. Furthermore, if only the contributions of relativistic particle species are considered, then $ \mathcal{E}_{\ast} = \mathcal{P}_{\ast} = g_{\ast} $. Consequently, as expected, $ p = \rho / 3 $ is recovered, corresponding to an expansion dominated by radiation (RD) and a single-fluid interpretation. By taking only into account relativistic particles, the time--temperature relation can be simply found by the substitution of Equation~\eqref{radiation thermal expansion rate} into Equation~\eqref{expansion rate non vacuum epochs}, which yields
\begin{equation}\label{radiation time temperature relation}
t=\frac{0.301}{\sqrt{g_{\ast}(T)}}\frac{m_{Pl}}{T^2}\sim \left(\frac{T}{\textrm{MeV}}\right)^{-2}~\textrm{sec}.
\end{equation}

Indeed, if we focus solely on the contributions from relativistic species and ignore how $\mathcal{E}$ changes over time—assuming $\mathcal{E} = \mathcal{P} = \mathcal{S} = g_{\ast}$ remains constant—the same result can be achieved. This can be confirmed by integrating the precise time--temperature relationship outlined in Equation~\eqref{exact time-temperature relation}.
In equilibrium, the temperature of the universe is expected to decrease over time according to the relation $t \propto T^{-2}$. Therefore, if equilibrium is maintained, the early universe would be characterized by extremely high temperatures and would be predominantly filled with relativistic particles ($T \gg m_i$). During this era, all interactions would be mediated by massless gauge bosons, with their cross sections described by $\sigma_k \sim \alpha^n T^{-2}$. Here, $\alpha$ represents the coupling constant relevant to the process, typically ranging from 0.01 to 0.1, and $n=1$ or $2$ corresponds to decays or two-body reactions, respectively.
Thus, in the high-temperature regime of the early universe, the scattering cross sections for various processes are approximately given by the same formula. This also applies to the interaction rates per particle as indicated by the same expression:
\begin{equation}
\Gamma\equiv n_T\sigma \vert v\vert\,,
\end{equation}
where $v$ represents the relative velocity between particles, and $n_T$ denotes the number density of target particles. Given that the universe is primarily composed of relativistic particles, where the velocity $|v|$ is approximately 1 and the number density $n_T$ scales with the cube of the temperature ($T^3$), interaction rates are generally estimated to be $\Gamma \sim \alpha^n T$.\linebreak Thus, requiring the thumb rule $\Gamma>H$ as a minimal condition for equilibrium allows to conclude that for
\begin{equation}
T>0.602\frac{\alpha^n m_{Pl}}{\sqrt{g_{\ast}}}\sim 10^{16}~\textrm{GeV}\label{thermal bound GR1} \, ,
\end{equation}
equilibrium may be established at high temperatures; however, below the threshold\linebreak defined by Equation~\eqref{thermal bound GR1}, all perturbative interactions are expected to freeze out,\linebreak rendering them ineffective in maintaining or establishing thermal equilibrium.\linebreak Paradoxically, the equilibrium condition is satisfied when the time is short\linebreak ($t > 10^{-38}$ s), or equivalently, when the temperature is high but constrained by the bound given in Equation~\eqref{thermal bound GR1}. At temperatures exceeding $10^{16}$ GeV or times earlier than $10^{-38}$ s, neither known interactions nor those predicted by Grand Unified Theories are capable of thermalizing the universe. 

However, the possibility of other processes capable of establishing thermal\linebreak equilibrium at temperatures below the bound given by Equation (\ref{thermal bound GR1}) cannot be excluded.\linebreak For instance, quantum gravitational processes occurring before the Planck time, $t_{Pl} \sim 10^{-43}$ s, might play a significant role~\cite{Fischetti:1979ue,Hartle:1980nn,Hartle:1980kz}. For temperatures above $~200$ GeV, all known species of elementary particles, that is, all particles of the SM, starting from the photon till the top quark ($m_t=172.69 \pm 0.30$) GeV~\cite{ParticleDataGroup:2022pth}, are relativistic. In fact, in the\linebreak SM scenario, particles only acquire mass after the electroweak (EW) phase transition such that for temperatures above the EW scale,  $\mathcal{E}_{\ast_{\textrm{SMP}}}=\mathcal{P}_{\ast_{\textrm{SMP}}}=\mathcal{S}_{\ast_{\textrm{SMP}}}=g_{\ast_{\textrm{SMP}}}=106.75$, and the previous description remains perfectly exact. It is evident that when annihilations or decays are not occurring, the aforementioned statement holds true. However, at temperatures below approximately 100 GeV, the masses of particle species become significant, causing their number densities, energy densities, and pressure to decline as the temperature approaches their mass thresholds. This reduction occurs through particle annihilation or, if the particles are unstable, through their decay. 

At higher temperatures, such reactions are continually occurring; however, they are balanced by particle--antiparticle pair production or inverse decay processes. In contrast,\linebreak at lower temperatures, the thermal energies of the particles are insufficient to sustain these balancing processes. Consequently, the contribution of a given particle species to the energy density and pressure diminishes gradually, rather than instantaneously, often requiring several Hubble times to become negligible. During periods when a particle species is semi-relativistic, where it is not fully described by either $T \ll m_i$ or $T \gg m_i$, its impact on total pressure, energy density, and entropy density can be significant. In these cases, $\mathcal{E}{\ast}$, $\mathcal{P}{\ast}$, and $\mathcal{S}{\ast}$ do not remain constant and must be calculated numerically. However, this scenario is relatively rare compared to the two primary cases: relativistic and non-relativistic.\linebreak Even when considering these intermediate cases, $\mathcal{E}{\ast}$ changes slowly, making it reasonable to approximate $\mathcal{E}{\ast} \approx g{\ast}$ (constant) and $\mathcal{E}{\ast} \approx \mathcal{P}{\ast}$. Thus, for most practical purposes regarding the expansion time scale, the approximate result given by Equation~\eqref{radiation time temperature relation}\linebreak is adequate.

However, for the relation between $a$ and $T$, a more accurate relation is needed and should indeed be provided by the result obtained by the conservation of the entropy and given by Equation~\eqref{entropy conservation scale factor temperature relation}. Indeed, for as long as the equilibrium situation is held and only the contributions from the relativistic particles are being accounted for, $\mathcal{S}_{\ast}\simeq \mathcal{E}_{\ast}$. This is of course false if the relativistic content is not entirely in equilibrium or if the contribution of semi-relativistic particles is being taken into account since in this last situation, $\mathcal{P}_{\ast}\neq \mathcal{E}_{\ast}\neq cte$ and thus $\mathcal{S}_{\ast}\neq \mathcal{E}_{\ast}\neq cte$ and it may be indeed more appropriate to use the accurate result given by Equation~\eqref{exact time-temperature relation}. Yet, the conservation of the entropy allows to establish an exact relation between the scale factor and temperature at two different times, skipping this kind of detail. 

Furthermore, as the temperature falls and some particles species become\linebreak non-relativistic, the reaction rates for the processes involving these species as target\linebreak particles also become exponentially suppressed. Additionally, massive bosons that\linebreak mediate some types of interactions may also become non-relativistic and the cross section for the interactions involving these bosons may also become suppressed by the boson rest masses ($\sigma \sim \alpha^2T^2/m_X^4$), thus suppressing the respective interaction rates as well.\linebreak Each of these scenarios, individually or combined, may disrupt equilibrium due to the interaction rates becoming insufficient ($\Gamma_{int}<H$). 

As a result, the rest masses that remain after annihilations (and possible decays) start to dominate, marking the transition of the universe into its matter-dominated (MD) epoch. This phase represents the universe's ``adolescence'' phase. This shift mirrors what happened with nucleons and electrons: as the temperature dropped below their rest masses, they annihilated with their antiparticles. For nucleons, this annihilation began right after their formation during the quantum chromodynamics (QCD) phase transition.\linebreak However, there was a slight excess of nucleons compared to antiparticles, allowing this\linebreak surplus to survive the annihilation. Since nucleons are the lightest baryons, the baryon\linebreak number today is mainly found in protons and neutrons. Therefore, for temperatures $T\lesssim 10$ MeV, the baryon number is effectively concentrated in nucleons
\begin{equation}
n_{\bar{N}}\ll n_N \quad \textrm{and} \quad n_N\equiv n_n+n_p=n_B\,,
\end{equation}
where $n_{\bar{N}}$, $n_{N}$, $n_n$, and $n_p$ are the number densities of nucleons, antinucleons, neutrons, and protons, respectively. On the other hand, since the universe is electrically neutral and the negative charge lies in the electrons, after the electron--positron annihilation, the number\linebreak of electrons must equal the number of protons.

\section{Out-of-Equilibrium Phenomena}\label{sec:Out-Of-Equilibrium Phenomena}

While the previous description captures the thermodynamic evolution of the early universe, a thorough understanding also requires considering out-of-equilibrium phenomena~\cite{Stewart:1969a}. Many aspects of new physics, such as dark matter, have speculative results that are built in out-of-equilibrium scenarios, processes like freeze-in, a pivotal process for dark-matter~\cite{Hall:2009bx,Chu:2011be,Hryczuk:2023bob,Bernal:2017kxu}, and freeze-out~\cite{Baer:2014eja,Baldes:2017gzw}. Both processes play a pivotal role in the full description of the early universe and in this review, we will give a brief introduction to the freeze-out mechanism. Therefore, in this section, we present  a comprehensive review of out-of-equilibrium phenomena in the context of the primordial universe. Once more, this section draws inspiration and some results from~\cite{Kolb:1990vq} and for more in-depth details of most of the topics presented, we recommend~\cite{Treciokas1971IsotropicSO,ellis1983anisotropic,Groot1980RelativisticKT,Gondolo:1990dk}.

\subsection{Boltzmann Equation in the Expanding Universe\label{subsec:Boltzmann Equation in the Expanding Universe}}

As previously noted, the Boltzmann equation is the formal tool to describe\linebreak  the evolution of scenarios beyond equilibrium, as it provides a crucial framework for accurately tracking particle phase space distributions. Its significance is further enhanced by the fact that, in most physically relevant scenarios, it can be approximately solved under well-justified assumptions. This yields detailed results regarding the present-day abundances of various species, thereby offering valuable insights into the thermal history of the universe and its potential implications for key areas of microphysics. In its general form, the Boltzmann equation in Hamiltonian formalism can be expressed as
\begin{equation}
\mathbf{\hat{L}}[f]=\mathbf{C}[f]\,,\label{boltzmann equation 1a}
\end{equation}
where $\mathbf{C}$ is the collision operator and $\mathbf{\hat{L}}$ is the Liouville operator. In its covariant form, the Liouville operator is given by~\cite{Treciokas1971IsotropicSO,ellis1983anisotropic}
\begin{equation}
\mathbf{\hat{L}}(f)=p^c\frac{\partial f}{\partial {x^c}}-\Gamma^{c}_{ab}\,p^ap^b \frac{\partial f}{\partial{p^c}}, 
\end{equation}
where $\Gamma^{c}_{ab}$ is the metric connection and $p^a=(E,\,\mathbf{p})$ is the quadri-momentum. For the Robertson--Walker metric, the non-zero components of the connection can easily be found (see for instance~\cite{Wald:1984rg,Weinberg:1972kfs,dInverno:1992gxs}) and since in this case $f(\mathbf{x},\,\mathbf{p},t)=f(\vert\mathbf{p}\vert,\,t)=f(E,\,t)$,\linebreak the Liouville operator may be cast in the form
\begin{equation}
\mathbf{\hat{L}}[f(E,~t)]=E\frac{\partial f}{\partial t}-H\vert\mathbf{p}\vert^2\frac{\partial f}{\partial E}\,.\label{Liouville operator}
\end{equation}

For a given particle species $i$, the Boltzmann Equation~\eqref{boltzmann equation 1a} can be written in terms of its density of states $dn_i$, using $f=(2\pi)^3dn_i/g_i\mathrm{d}^3p_i$ so that upon integration by parts, it can be cast in the form (see for instance~\cite{Kolb:1990vq} for further details)
\begin{equation}
\frac{dn_i}{dt}+3Hn_i=I_i^{\textrm{coll}}\,,\label{formalBoltzmannGR1}
\end{equation}
where
\begin{equation}
I_i^{\textrm{coll}}\equiv g_i/(2\pi)^3\int\mathbf{C}[f_i]\mathrm{d}^3p_i \, ,
\end{equation}
is the collision integral, which for a generic process of the type $i_1+i_2+...+i_n\leftrightarrow j_1+j_2+...+j_n$, is given by
\begin{align}
I_{i_1}^{\textrm{coll}}=-&\int\,\left[W(i_1i_2...i_n\rightarrow j_1j_2...j_n)\prod^n_{k=1}f_{i_k}(1\pm f_{j_k})
 \right.
 	\\ \nonumber
&\left.-\,W(j_1j_2...j_n\rightarrow i_1i_2...i_n)\prod^n_{k=1}f_{j_k}(1\pm f_{i_k})\right]\,\mathrm{d}\Pi_{i_k}\mathrm{d}\Pi_{j_k}\,.
\end{align}
$f_{a_k}$ is the phase space densities for any species $a_k$ and the sign ($+$) applies to boson species and the sign ($-$) to fermions species and
\begin{equation}
W(i_1i_2...i_n\rightarrow j_1j_2...j_n)\equiv(2\pi)^4\delta^4\left(\sum_{k=1}^n p_{i_k}-\sum_{k=1}^np_{j_k}\right)
\vert\mathcal{M}(i_1i_2...i_n\rightarrow j_1j_2...j_n)\vert^2
\end{equation}
with $\vert\mathcal{M}(i_1i_2...i_n\rightarrow j_1j_2...j_n)\vert^2$, that sometimes is called the Feynman amplitude, being the matrix element squared for the generic process $\alpha+\beta...\rightarrow\lambda+\sigma$ averaged over the initial and final spins and including the appropriate symmetry factors for identical particles in the initial or final states (we refer to~\cite{Quigg+2014} for how to calculate this term). 
Finally, $d\Pi_{i_k}$ \linebreak is given by
\begin{equation}
d\Pi_{i_k}\equiv \frac{g_{i_k}}{(2\pi)^3}\frac{\mathrm{d}^3p_{i_k}}{2E_{i_k}}\,,
\end{equation}
where $g$ counts the internal degrees of freedom. If a particle species is involved in more than one process, it is essential to include all of these processes in the collision term.

The Boltzmann Equation \eqref{formalBoltzmannGR1} is a powerful tool of statistical mechanics, as it provides a complete statistical description of interactions (for an in-depth discussion of this topic, we refer the reader to~\cite{Bernstein:1988bw}). In general, the Boltzmann equations constitute a coupled system of integral--differential equations governing the phase space distributions of all species present. Nonetheless, in the context of solving practical problems, the phase space distribution functions of all species, except for a few, can be approximated by their equilibrium distributions due to their rapid interactions with other species. Consequently, this simplification reduces the problem to a single integral--differential equation for the species of interest, which we denote in the following work as $\chi$. 

Furthermore, in the scenario where $\chi$ represents a stable particle, and assuming that all other species maintain equilibrium phase space distributions, the entropy per comoving volume remains conserved. This approximation holds because, with the exception of the species $\chi$, all other species are assumed to have negligible chemical potentials and maintain equilibrium distribution functions. Therefore, in this case, it is useful to introduce the quantity $X=n_\chi/s$ as a dependent variable, which is related to the actual number of $\chi$'s in a comoving volume and allows to write the left-hand side (LHS) of Equation \eqref{formalBoltzmannGR1} as~\cite{Kolb:1990vq}
\begin{equation}
\vspace{-6pt}
\dot{n}_{\chi}+3Hn_{\chi}=s\dot{X} \, .
\end{equation}

In the same way as performed in the {Section} \ref{subsec:The matter content of the Universe}, it is useful to introduce the adimensional variable that is the ratio of the mass of $\chi$ and temperature: $z=m_\chi/T$. 

Considering the subject of the following sections, it is useful to introduce a hypothetical cosmological solution in the form of a power law. This solution is derived from the field equations of a theory that, while not necessarily GR, is still bound by the requirement to satisfy the covariant conservation of energy \eqref{Conservation}. This solution is characterized\linebreak by the following:
\begin{align}
\rho(t)&=\frac{P_i}{a^{3\gamma_i}(t)}\label{hypothetical energy density}\, ,\\
\frac{a(t)}{A_i}&=(t-t_{i0})^b, \qquad H(t)=\frac{b}{t-t_{i0}}\label{hyptothetical}\,,
\end{align}
where $t_{i0}$, $A_i$, and $P_i$ are constants of integration, and $b$ is a constant parameter. As clearly demonstrated by Equation \eqref{hypothetical energy density}, this hypothetical solution obeys the conservation of energy. Consequently, the conservation of entropy per comoving volume is maintained under the same conditions as those prescribed by GR. In fact, the GR cosmological solutions, described by Equation~\eqref{expansion rate non vacuum epochs}, are recovered when $b = 2/3\gamma_i$. Therefore, the time--temperature relation for this solution can be succinctly expressed as follows:
\begin{equation}
t-t_{i0}=C_iT^{-1/b}\,,\label{hypothetical time-temperature relation}
\end{equation}
which yields the following differential relation:
\begin{equation}
dt=\frac{C_i}{b} m_\chi^{-1/b} z^{\frac{1-b}{b}}dz=\frac{dz}{z H(z)}=\frac{z^{\frac{1-b}{b}}}{H(z=1)}\,dz\,,\label{differential z}
\end{equation}
with $C_i$ being a constant which can be easily determined by specifying $b$ and the thermal expansion rate, $H(T)$, namely through the relation $H(z=1)=m_\chi^{1/b}H(T=1)=bm_\chi^{1/b}/C_i$. Note that a factor $\mathcal{S}_{\ast}$ should indeed have been included in Equation~\eqref{hypothetical time-temperature relation}.\linebreak Consequently, an additional term involving its derivative should have been incorporated into Equation~\eqref{differential z}. However, this latter term is small and, to a good approximation, $\mathcal{S}_{\ast}$\linebreak can be fixed by the value of $g_{\ast S}$ around the freeze-out. Additionally, there are\linebreak well-motivated approximations that significantly simplify the collision integral, the first of which is the use of Maxwell--Boltzmann statistics for all species instead of the exact statistics. This assumption neglects the scenarios of Bose--Einstein condensation or \mbox{Fermi degeneracy.} In their absence, the blocking and stimulated emission factors can be disregarded, as $1 \pm f \simeq 1$~\cite{Kolb:1990vq}. Another simplification can be established by assuming CP invariance\linebreak which implies
\begin{equation}
    \vert\mathcal{M}(i_1i_2...i_n...\rightarrow j_1j_2...j_n...)\vert^2=\vert\mathcal{M}(j_1j_2...j_n\rightarrow i_1 i_2...i_n)\vert^2 = \vert\mathcal{M}\vert^2 \, .
\end{equation}

Under these two assumptions, the collision integral becomes rather reduced and simplified  and adopting the hypothetical cosmological solution (\ref{hypothetical energy density}) and (\ref{hyptothetical}), the Boltzmann Equation \eqref{formalBoltzmannGR1} assumes the simple form
\begin{equation}
X'=-\frac{z^{\frac{1-b}{b}}}{s(z)H(z=1)}\int W(i_1i_2...i_n\rightarrow j_1j_2...j_n)\prod_{k=1}^n[f_{i_k}-f_{j_k}]\,\mathrm{d}\Pi_{i_k}\mathrm{d}\Pi_{j_k}\,,
\end{equation} 
where the prime $'$ stands for $d/dz$, and $s(z)=2\pi^2g_{\ast S}m_\chi^3z^{-3}/45=s(z=1)z^{-3}$ is the entropy density written in terms of the variable $z$.

\subsection{Stable Particles: The Freeze-Out of Species\label{subsec:Stable Particles: The Freezing of Species}}

The kinetic Equation \eqref{formalBoltzmannGR1}, which governs the evolution of the number density of stable particles (particles with lifetimes significantly 
 longer than the age of the universe); besides stable particles, one can also consider unstable particles, bringing new considerations such as if a given massive particle species $\lambda$ is unstable but relatively long lived, it can decouple from the surrounding medium before decaying (for further details see~\cite{Scherrer:1984fd})---we will not explore this case in this work, rather leaving it for further studies) can be solved approximately with a high degree of accuracy by introducing additional assumptions beyond those previously considered. These supplementary assumptions can be summarized as follows:

\begin{enumerate}

\item \emph{\textbf{Only annihilations and inverse annihilation processes need to be included}}. 

Given that the particles are stable, it is generally adequate to consider only annihilation and inverse annihilation processes when tracking the evolution of their phase space distribution. These processes are primarily responsible for any significant changes in the number of particles of interest~\cite{Kolb:1990vq}. We are considering processes such as
\begin{equation}
    \chi \bar{\chi} \leftrightarrow K \bar{K} \, ,
\end{equation}
with $K$, $\bar{K}$ standing for all the species into which $\chi$'s can annihilate. For simplicity, we only consider 2vs2 annihilation and that the asymmetry between $\chi$'s and $\bar{\chi}$'s is negligible. In this scenario, we consider that the particle species $\bar{K}$, $\bar{K}$ are in complete equilibrium and their chemical potentials are zero. Conversely, if chemical equilibrium is not maintained, the chemical potentials for particles and antiparticles are equal in magnitude but opposite in sign such that $\mu_\chi = -\mu_{\bar{\chi}}$.

\item \emph{\textbf{The products of annihilation are in complete thermal equilibrium.}}
 
These products generally experience interactions that are stronger than those between the original ``parent'' particles~\cite{Kolb:1990vq}. As a result, they will rapidly achieve thermal equilibrium after their formation.

\item \emph{\textbf{The particle species under inspection is in kinetic equilibrium}}. 

In this scenario, its distribution function has the form~\cite{Kolb:1990vq}
\begin{equation}
f_\chi(p)=e^{-(E-\mu)/T}=e^{\mu/T}f_{\chi}^{\textrm{eq}}(p)\,,\label{integral distribution1}
\end{equation}
so that $\exp(\mu/T) = n_\chi / n_{\chi}^{\text{eq}}$, where $n_\chi$ represents the actual number density and $n_{\chi}^{\text{eq}}$ is the equilibrium number density. This assumption is justified by the fact that as the number density of the relevant particle species decreases, the annihilation of these particles requires a partner species with a similarly suppressed number density. Since chemical equilibrium is maintained through annihilation processes, it will be disrupted sooner than the kinetic equilibrium, which only requires interactions with abundant massless particles. Given that the particle species is stable and decay processes are excluded, the collision integral can be expressed as $I^{\text{coll}} = I^{\text{el}} + I^{\text{ann}}$, where $I^{\text{el}}$ and $I^{\text{ann}}$ denote the collision integrals for elastic scattering and annihilation, respectively. For non-relativistic particles (where $m > T$), the elastic scattering integral $I^{\text{el}}$ is significantly larger than the annihilation integral $I^{\text{ann}}$ due to the exponential suppression of heavy particles. Consequently, the large value of $I^{\text{el}}$ enforces kinetic equilibrium. Therefore, for a distribution function of the form given by \eqref{integral distribution1}, integrating the Boltzmann equation over $d\Pi$ results in the disappearance of this large integral, though its effect is still reflected in the distribution function \eqref{integral distribution1}~\cite{A.D. Dolgov}.\\

\end{enumerate}

Under these assumptions, the collision integral for a particle species $\chi$ following the general process $\chi\bar{\chi}\rightarrow j_1 \bar{j_1}$ can be integrated, yielding~\cite{Kolb:1990vq}
\begin{equation}
\vspace{-6pt}
    I_{\chi}^{\textrm{coll}}=\left\langle\sigma (\chi\bar{\chi}\rightarrow j_1\bar{j_1})\vert v\vert\right\rangle[(n_{\chi}^{\textrm{eq}})^2-n_{\chi}^2]
\end{equation}
where $\left\langle\sigma (\chi\bar{\chi}\rightarrow j_1\bar{j_1})\vert v\vert\right\rangle$ is the thermally averaged annihilation cross section times the\linebreak relative velocity, and the Boltzmann equation resumes then to the form of the\linebreak Zeldovich--Lee--Weinberg equation, which for any hypothetical cosmological solution of the type given by Equations \eqref{hypothetical energy density} and \eqref{hyptothetical} can be written as (for further details see~\cite{Kolb:1990vq}))

\phantom{0}\vspace{-12pt}
\begin{align}
z\frac{X'}{{X^{\textrm{eq}}}}&=\frac{\Gamma_A}{H(z)}\left[1-\left(\frac{X}{X^{\textrm{eq}}}\right)^2\right]\label{Lee Weinberg 1}\,,
\end{align}
where $\Gamma_A \equiv n_{\chi}^{\textrm{eq}}\left\langle\sigma_A\vert v\vert\right\rangle=sX^{\textrm{eq}}\left\langle\sigma_A\vert v\vert\right\rangle$ is the \emph{total} annihilation cross section; $\sigma_A$ is the thermally averaged \emph{total} scattering cross section, which is defined as including all possible final states and thus all possible annihilation channels; and ${X}^{\textrm{eq}}$ is the equilibrium value \mbox{of $n_{\chi}/s$.} 

From Equation \eqref{Lee Weinberg 1}, it is evident that the ratio $\Gamma/H$ serves as a quantitative indicator of the effectiveness of the processes under consideration. Specifically, as $\Gamma/H$ approaches values significantly less than unity, the relative change in the number of $X$ particles within a comoving volume diminishes. Consequently, as $\Gamma/H$ decreases with time or temperature, the efficiency of the annihilation processes wanes. This decline in effectiveness results in an increasing deviation of the actual number of $\chi$ particles in a comoving volume from its equilibrium value. Eventually, the annihilation processes become negligible, leading to a state where the number of $\chi$ particles effectively ``freezes in'' within the comoving volume. This observation provides a novel perspective on the implications of the thumb rule (\ref{coupled}) and (\ref{decoupled}) and the freezing relation (\ref{freezing relation1}). 

Equation (\ref{Lee Weinberg 1}) can be solved approximately~\cite{Kolb:1990vq} by making some simple assumptions, which involve considering two distinct temperature regimes. To achieve this, it is useful to parameterize the temperature dependence of the annihilation cross section, which can generally be performed in a manner similar to that used in~\cite{Kolb:1990vq}, by
\begin{equation}
\left\langle \sigma_A\vert v\vert \right\rangle =\sigma_{0} z^{-n}(1+dz^{-m})\,,
\end{equation}
where $n$, $m$, and $d$ are constant parameters depending on the type of annihilation by which the processes proceed. Using this parametrization, Equation~\eqref{Lee Weinberg 1} can be written as
\begin{equation}
\frac{X'}{(X^{\textrm{eq}})^2}=-\lambda z^{(-n+1/b-4)}(1+dz^{-m})\left[1-\left(\frac{X}{X^{\textrm{eq}}}\right)^2\right]\,,\label{Lee Weinberg2}
\end{equation}
where
\begin{equation}
\lambda=\sigma_0 \left(\frac{s}{H}\right)_{z=1}\, ,
\end{equation}
with $s(z=1) = \frac{2\pi^2}{45}g_{\ast S} m_\chi^3$. Thus, assuming that at high temperatures (low $z$), $X$ follows very near its equilibrium value, that is, $X$ is weakly deviated from the equilibrium ($X\approx X^{\textrm{eq}}$), it can be written that $X=X^{\textrm{eq}}(1+\delta X)$, where $\delta X\ll 1$ and since in this situation both $X^{\textrm{eq}}$ and $X$ are varying very slowly and approximately by the same amount, then $(X^{\textrm{eq}})'\approx X'$. Thus, neglecting the quadratic terms in $\delta X$, the solution to Equation~\eqref{Lee Weinberg2} is given in this approximation by~\cite{A.D. Dolgov}
\begin{equation}
\delta X\approx -(2\lambda)^{-1}\,\left(\frac{z^{(n-1/b+4)}}{1+dz^{-m}}\right)\,\frac{(X^{\textrm{eq}})'}{(X^{\textrm{eq}})^2}
=(2\lambda X^{\textrm{eq}})^{-1}\,\left(\frac{z^{(n-1/b+4)}}{1+dz^{-m}}\right)\,\frac{K_1(z)}{K_2(z)} \, , \label{deviation GR1}
\end{equation}
where, in performing the last step, it is noticed that $X^{\textrm{eq}}\propto z^2 K_2(z)$ and $\frac{d}{dz}( z^2K_2(z))=-z^2 K_{1}(z)$. As the universe continues to expand and cool, it will eventually reach a critical redshift, $z_{\ast}$, where the deviation from equilibrium becomes significant, reaching a value of unity. Beyond this point, Equation~\eqref{Lee Weinberg2} can be approximated as
\begin{equation}\label{Lee Weinberg asymptotic2}
X'\approx - \lambda z^{(-n-4+1/b)}(1+dz^{-m})X^2 \, .
\end{equation}

If we assume that this regime is initiated by the condition $\delta X = k$, where $k$ is a constant on the order of unity (with $k = 1$ generally providing a reasonably accurate result), then Equation~\eqref{deviation GR1} can be used to derive an expression for a threshold redshift, denoted as $z_{\ast}$. Above this threshold, the deviation from equilibrium becomes significant, and one has
\phantom{0}\vspace{-12pt}
\begin{equation}\label{z ast}
z_{\ast}^{(n-1/b+4)}
=2\lambda k (1+dz_{\ast}^{-m})X^{\textrm{eq}}(z_{\ast})\frac{K_2(z_{	\ast})}{K_1(z_{\ast})}\,.
\end{equation}

It thus becomes evident that $z_{\ast}$ provides a refined estimate for the redshift at which freeze-out actually occurs. However, in most of the cases, there is typically a small deviation from $z_\ast$ from the actual value at which the freezing occurs, $z_f$. Additional care must be taken when interpreting the true meaning of $z_{\ast}$ and $z_f$ as depending on the exponent of $z$, as it is possible that $\delta X$ or $\Gamma_A/H$ may not necessarily increase (or decrease) with $z$. In such cases, as $z$ increases, $X$ may approach its equilibrium value, and $z_{\ast}$ or $z_f$ would then indicate the point where equilibrium is established rather than the point of freeze-out. However, within the context of GR, this scenario typically does not occur, as equilibrium is established very early on. Consequently, this consideration will not be explored \mbox{further here}. 

Considering $1/b\neq n+m+4$ and $1/b\neq n+4$, the integration of Equation~\eqref{Lee Weinberg asymptotic2} from $z_{\ast}$ to $z$ gives
\begin{align}
X(z)&=\frac{B_1}{\frac{B_1}{X(z_{\ast})}+\xi}
\,,\label{Boltzmann solution GR1}
\end{align}
where $B_1=(n-1/b+3)(m+n-1/b+3)\lambda^{-1}$ and $\xi$ being given by 
\begin{eqnarray}
    \xi &=& \left[\left(-n+\frac{1}{b}-3\right)(1+dz^{-m})-m\right]z^{(-n+\frac{1}{b}-3)}
	\nonumber \\    
    && +\left[\left(n-\frac{1}{b}+3\right)(1+dz_{\ast}^{-m})+m\right] z_{\ast}^{(-n+\frac{1}{b}-3)} \, ,
\end{eqnarray}

The case where $1/b=n+4$ and $1/b\neq n+m+4$, the integration yields
\begin{equation}
X(z)=\left[\lambda\ln\left(\frac{z}{z_{\ast}}\right)+\frac{d(z^{-m}-z_{\ast}^{-m})}{m+1}+\frac{1}{X(z_{\ast})}\right]^{-1}\,.\label{Boltzmann solution GR2}
\end{equation}

In the case of an isentropic expansion, the results from Equations (\ref{Boltzmann solution GR1})--(\ref{Boltzmann solution GR2}),\linebreak combined with Equation~\eqref{z ast} offer a method to approximately track the evolution of the actual value of $X$ in the temperature range where $z > z_{\ast}$. In this context, the final abundance of $X$, or its present-day abundance, is simply given by the asymptotic value $X_f \equiv X(z \rightarrow \infty)$.\linebreak Assuming that the conditions $1/b - n - 3 < 0$ and $1/b - n - 3 - m < 0$ hold, which \linebreak is generally true within GR, the final abundance is provided by Equation (\ref{Boltzmann solution GR1}) giving
\begin{equation}\label{final saturation value annihilation}
X_f=\frac{\left(n-\frac{1}{b}+3\right)z_{\ast}^{n-\frac{1}{b}+3}}{\lambda\left\{1+\left(\frac{n-\frac{1}{b}+3}{m+n-\frac{1}{b}+3}\right)dz_{\ast}^{-m}+\left[\frac{2k(n-\frac{1}{b}+3)(1+dz_{\ast}^{-m})}{(1+k)z_{\ast}}\right]\,\frac{K_2(z_{\ast})}{K_1(z_{\ast})}\right\}}\,,
\end{equation}
where $X(z_{\ast})=(1+\delta X(z_{\ast}))X^{\textrm{eq}}(z_{\ast})=(1+k)X^{\textrm{eq}}(z_{\ast})$. Equations (\ref{z ast}) and (\ref{final saturation value annihilation}) provide a powerful and simple tool to treat the freeze out, both of relativistic and non-relativistic particle species, with the means provided by the asymptotic forms of the $K_n(z)$ functions.

\subsubsection{Relativistic Freezing: Hot Relics}

Considering that the deviation from the equilibrium becomes large when the particle species $X$ is still relativistic ($T\gg m$), then one can consider, in good approximation, $K_1(z)/K_2(z)\simeq z/2$ and so, the substitution of Equation~\eqref{z ast} into Equation~\eqref{final saturation value annihilation} leads to
\begin{align}
X_f&\simeq X^{\textrm{eq}}(z_{\ast})\left\{\frac{1}{k+1}+\left[\frac{(n-\frac{1}{b}+3)+m/(1+dz_{\ast}^{-m})}{4c(m+n-\frac{1}{b}+3)(n-\frac{1}{b}+3))}\right]z_{\ast}^2\right\}^{-1} \, 
	\nonumber \\
&\approx (k+1)X^{\textrm{eq}}(z_{\ast})\simeq \frac{45 g_{\chi} (k+1)}{2\pi^4 g_{\ast S}(z_{\ast})}\, ,
\end{align}
where it is assumed in the last steps that $z_{\ast}^2<4\lambda B_1$. In this case, due to the differences in the MB statistics relative to the exact ones, $g_\chi$ should be exchanged by $\zeta(3)g_\chi^n$. 

This correction reduces the error associated with using Maxwell--Boltzmann statistics and results in 
\begin{equation}
    X_f = \frac{45 \zeta(3) g_\chi^n (k+1)}{2\pi^4 g_{\ast S}(z_{\ast})} \, ,
\end{equation}
which is evidently equivalent to $ n^{\textrm{eq}}(z_{\ast})/s(z_{\ast}) $, where $ n^{\textrm{eq}} $ represents the equilibrium number density calculated using the exact distribution functions in the relativistic limit. Therefore, the freeze-out of relativistic species is insensitive to details of the freeze-out (decoupling) since the value of $X_f$ is just the equilibrium value of $X$ at the moment it freezes in and thus $z_{\ast}$ is  manifested only on the adequate value of $g_{\ast S}$, which should be used. The freezing of relativistic species can in good approximation  be computed by the \mbox{freezing relation.}

\subsubsection{Non-Relativistic Freezing: Cold Relics}

Considering that the freezing occurs when the species are already non-relativistic ($m_{\chi}\gg T$) requires special attention to the details of the freeze-out. Since in this case $z_{\ast}\gg1$, then $K_{1}(z_{\ast})/K_2(z_{\ast})\simeq 1$, which, after substitution in Equations~\eqref{z ast} and \eqref{final saturation value annihilation}, respectively results in
\begin{eqnarray}
z_{\ast} & \simeq &\left[\frac{2\lambda k (1+dz_{\ast}^{-m})}{z_{\ast}}X^{\textrm{eq}}(z_{\ast})\right]^{\frac{1}{n-1/b+3}}
	\nonumber \\
&\simeq & \left[\frac{45}{2\pi^4}\sqrt{\frac{\pi}{8}}\frac{g_\chi }{g_{\ast S}}\lambda k(1+dz_{\ast}^{-m})z_{\ast}^{1/2}e^{-z_{\ast}}\right]^{\frac{1}{n-1/b+3}}\,,
\label{general zf1}
\end{eqnarray}
\begin{equation}
X_f\simeq \frac{(n-1/b+3)z_{\ast}^{(n-1/b+3)}}{\lambda\left[1+\left(\frac{n-1/b+3}{m+n-1/b+3}\right)dz_{\ast}^{-m}+\frac{2k(n-1/b+3)(1+dz_{\ast}^{-m})}{(1+k)z_{\ast}}\right]}\label{general yf1}\,.
\end{equation}

In contrast, the equilibrium scenario gives
\begin{equation}
    X^{eq}=\frac{45}{2\pi^4} \sqrt{\frac{\pi}{8}} \frac{g}{g_{\ast S}} z^{3/2} e^{-z} \, ,
\end{equation}

The value of $z_{\ast}$ can be computed numerically or it can be given semi-analytically by
\begin{eqnarray}
z_{\ast}&\approx & \ln\left(\sqrt{\frac{2}{\pi^3}}\frac{g_{\chi}}{s(z=1)}\lambda \right)-\left(n+\frac{5}{2}-\frac{1}{b}\right)\,\ln z_{\ast}+\ln(1+dz_{\ast}^{-m})
	\nonumber\\
&\simeq & \ln\left(\sqrt{\frac{2}{\pi^3}}2\lambda\right)-\left(n+\frac{5}{2}-\frac{1}{b}\right)\,\ln \left[\ln\left(\sqrt{\frac{2}{\pi^3}}\frac{g_{\chi}}{s(z=1)}\lambda \right)\right] 
	\nonumber\\
&&+ \ln\left\{1+d\left[\ln\left(\sqrt{\frac{2}{\pi^3}}\frac{g_{\chi}}{s(z=1)}\lambda \right)\right]^{-m}\right\}\,,
\end{eqnarray}
provided that the logarithmic term does not dominate. Using the GR solution for the radiation-dominated case, in which $H(z=1)=1.67g_{\ast}^{1/2}m_\chi^2/m_{Pl}$ and $1/b=2$ and so, $\lambda =0.262 (g_{\ast S}/g_{\ast}^{1/2})m_{Pl}m_\chi\sigma_0$, it is straightforward to observe that Equation~\eqref{general yf1}\linebreak results in
\begin{equation}
\vspace{-12pt}
X_f\simeq\left[\frac{3.80g_{\ast}^{1/2}(n+1)z_{\ast}^{(n+1)}}{g_{\ast S}m_{Pl}m_\chi\sigma_0}\right]\Bigg/\left[1+\left(\frac{n+1}{m+n+1}\right)dz_{\ast}^{-m}\right]\,,
\end{equation}
that for $m=0$, resulting in the same parametrization for $\left\langle \sigma_A\vert v\vert \right\rangle$ done in~\cite{Kolb:1990vq}, gives 
\begin{equation}
\vspace{-12pt}
X_f\simeq\left[\frac{3.80g_{\ast}^{1/2}(n+1)z_{\ast}^{(n+1)}}{g_{\ast S}m_{Pl}m_\chi\sigma_0}\right] \, ,
\end{equation}
which  coincides with the result obtained in~\cite{Kolb:1990vq}.

\section{Inflation}\label{sec:Inflation}

Originally introduced by Alan Guth in 1983~\cite{Guth:1980zm} as a solution to the flatness and horizon problems associated with the standard Big Bang cosmology~\cite{Dicke:1979jwq}, the concept of inflation was also proposed to address the monopole problem that arises from the spontaneous symmetry breaking (SSB) in Grand Unified Theories (see~\cite{Kolb:1990vq} for more details about these problems). As previously discussed, these SSB events would have triggered phase transitions, leading to the formation of topological defects. Monopoles are an example of such defects, and their overproduction during non-inflationary phase transitions results in a relic monopole abundance that is inconsistent with the observational data. In Guth's model, known as \emph{old inflation}, a brief period of exponential expansion is driven by a strong first-order phase transition. This implies that while inflation could potentially resolve the monopole problem associated with the SSB of GUTs, the phase transition triggered by the SSB could, in turn, provide the mechanism for generating inflation itself.

Generally, these phase transitions can be described as being driven by a scalar field\linebreak or a set of fields, commonly referred to as the inflaton (denoted by $\phi$). The evolution of the inflaton is determined by its finite-temperature effective potential, which initially has\linebreak a single minimum but develops a second minimum as the temperature decreases.\linebreak At a critical temperature, $T_{cr}$, these two minima become degenerate. As the effective finite-temperature potential evolves, the expectation value of the inflaton shifts from\linebreak a metastable (false) phase to a stable (true) phase. At temperatures significantly above the critical temperature, the metastable phase is entropically favored, possessing the lowest free energy density. However, as the temperature drops below the critical temperature, the free energy of the stable phase falls below that of the metastable phase. Nevertheless, the universe remains trapped in the false vacuum due to an energy barrier separating the two phases. In this scenario, as the universe undergoes supercooling in the false vacuum state, the energy density of the false vacuum becomes the dominant component of the universe's total energy density. This effectively acts like a cosmological constant, driving a period of exponential expansion. However, the challenge with this model is that while sufficient inflation is needed to solve the cosmological puzzles, there is no mechanism within this framework to transition out of the false vacuum state. Although this work was ground-breaking, this original mechanism specifically fell into oblivion with the\linebreak emergence of mechanisms with fewer problems but that kept the core idea of the old inflation, an extremely fast expansion of the universe. In recent years, it has become more widely accepted that inflation in the early universe does not occur through a strong\linebreak first-order phase transition. Instead, the most promising models are now based on the general principle of slow-roll inflation. Notable examples of these models include \textit{new inflation} or \textit{slow-roll inflation} (Linde, 1982~\cite{Linde:1981mu}; Albrecht and Steinhardt, 1982~\cite{Albrecht:1982wi}) and chaotic inflation (Linde, 1983~\cite{Linde:1983gd}). In these frameworks, the universe evolves smoothly from a false vacuum state to a true vacuum state. This transition occurs as the inflaton field gradually rolls from its initial false vacuum expectation value down an effective potential slope towards its true vacuum expectation value, thereby avoiding quantum tunneling. Inflation is a complex subject and we refer the reader to~\cite{Kolb:1990vq,Freese:1990rb,Linde:1990flp,Liddle:1992wi,Lyth:1998xn,Liddle:1999mq,Tsujikawa:2003jp,Bassett:2005xm,Guth:2007ng,Weinberg:2008hq,Baumann:2014nda,Baumann:2022mni} for more in-depth details about this concept and its sub-topics. This section is based on and was inspired by~\cite{Kolb:1990vq,Liddle:1999mq,Baumann:2022mni}. 

\subsection{Slow-Roll Inflation: The Mechanism and Its Dynamics}

To obtain inflation, one needs to ensure that the scale factor grows at an accelerating rate, $\Ddot{a}>0$. Attending this demand and using the acceleration Equation~\eqref{aceleration} in terms of the scale factor (neglecting $\Lambda$) 
\begin{equation}
    \frac{\Ddot{a}}{a} = -\frac{4\pi G}{3} (\rho + 3p) \, ,
\end{equation}
one can see that for $\Ddot{a}>0$, then $p > -\rho/3$, indicating a type of matter with negative pressure. This corresponds to a form of matter that induces a repulsive gravitational effect.\linebreak Such behavior is present in the vacuum energy, which leads to the exponential growth of the scale factor as described by the de Sitter universe solution in general relativity, seen in Equation \eqref{deSitterGR}. This rapid expansion is the hallmark of inflation.

From a more dynamic perspective, scalar fields also exhibit negative pressure, making them crucial components in the inflationary paradigm. Given this, it becomes evident that inflation is closely tied to the dynamics of scalar fields. To gain a deeper understanding of the slow-roll mechanism, it is essential to begin with the study of scalar field dynamics:
\begin{equation}\label{action scalar fied general}
    S_\phi = \int \mathrm{d}^4 x \sqrt{-g} \left[ -\frac{1}{2} g^{\mu\nu} \partial_\mu \phi \partial_\nu \phi - V(\phi) \right]\, ,
\end{equation}
where by employing the variation in the inverse metric gives the energy momentum of the scalar field
\begin{equation}
    T_{\mu\nu} = \partial_\mu \phi \partial_\nu \phi - \frac{1}{2}g_{\mu\nu}\left(g^{\alpha\beta}\partial_\alpha \phi \partial_\beta \phi + 2V(\phi) \right)\, ,
\end{equation}
allowing to compute the energy density and pressure as 
\begin{equation}\label{energy density scalar general}
   \rho_\phi=T_{\mu \nu} u^\mu u^\nu = \frac{1}{2} (u^\mu \nabla_\mu \phi)^2 + V(\phi) +\frac{1}{2} \vec{\nabla} \phi \cdot \vec{\nabla} \phi \, ,
\end{equation}
\begin{equation}\label{pressure scalar general}
   P_\phi = \frac{1}{3} T_{\mu\nu} h^{\mu\nu} = \frac{1}{2} (u^\mu \nabla_\mu \phi)^2 - V(\phi) +\frac{1}{6} \vec{\nabla} \phi \cdot \vec{\nabla} \phi \, ,
\end{equation}
which in turn, for a homogeneous field configuration, $\phi=\phi(t)$, gives 
\begin{equation}\label{energy density scalar}
   T_{00}=\rho_\phi = \frac{1}{2} \dot{\phi}^2 + V(\phi) \, ,
\end{equation}
\begin{equation}\label{pressure scalar}
   T_{11}=P_\phi = \frac{1}{2} \dot{\phi}^2 - V(\phi) \, .
\end{equation}

Additionally, using a flat FLRW metric and considering a homogeneous field\linebreak configuration, the action \eqref{action scalar fied general} reduces to
\begin{equation}
    S_\phi = \int \mathrm{d}t \mathrm{d}^3x \ a^3(t) \left[ \frac{1}{2} \dot{\phi}^2 - V(\phi) \right] \, ,
\end{equation}  
that by varying with respect to the scalar field, from the principle of least action, it gives the field equation
\begin{equation}\label{Klein-Gordon eq}
    \Ddot{\phi} + 3H\dot{\phi} = -V_\phi \, .
\end{equation}

This formulation corresponds to the Klein--Gordon equation, where subscript notation denotes differentiation with respect to the specified variable. From this equation, the term $3H\dot{\phi}$ is particularly significant, as it plays a crucial role in the development of slow-roll inflation. By taking the time derivative of the energy density, as defined in Equation \eqref{energy density scalar}, it is possible to derive the continuity equation and, when this is combined with the pressure Equation \eqref{pressure scalar} and the Klein--Gordon equation, it results in the standard form typically expected for the continuity equation
\begin{equation}
    \dot{\rho}_\phi = -3H(\rho_\phi + P_\phi)\, , 
\end{equation}
giving $\Ddot{a} \propto -(\rho_\phi + 3P_\phi)$. It is simple to see that the density and pressure are, in general,\linebreak not related by a conventional constant equation of state as the typical description for a perfect fluid. However, the main behavior of this relation is encoded in the relation between the kinetic term of the inflaton and its potential energy. If the kinetic term if much smaller than the potential energy, then $\rho_\phi \approx -P_\phi$. The inflaton potential acts as an temporary cosmological constant, being the source of the exponential expansion (behavior similar\linebreak to Equation \eqref{deSitterGR}). 

With this defined, the dynamics of inflation can be determined by combining the Friedmann equation 
\begin{equation}\label{Friemdann scalar}
    H^2 = \frac{1}{M_{Pl}^2}\left[ \frac{1}{2} \dot{\phi}^2 + V(\phi)\right] \, ,
\end{equation}
and the Klein--Gordon Equation \eqref{Klein-Gordon eq}, making a set of coupled equations. This coupling manifests itself as follows: the energy stored in the inflaton field influences and determines the Hubble rate, leading to a dynamic evolution of the friction term $3H\dot{\phi}$. This friction term, in turn, affects the evolution of the field itself, creating a feedback loop between the field dynamics and the expansion rate of the universe. Combining Equations \eqref{Friemdann scalar} and  \eqref{Klein-Gordon eq}, the time evolution of the Hubble parameter is given by
\begin{equation}\label{H scalar 1}
    \dot{H} = -\frac{1}{2} \frac{\dot{\phi}}{M^2_{Pl}} \, .
\end{equation}

Using this equation in combination with the first Hubble slow-roll parameter, $\varepsilon$,\linebreak an important parameter for the concretization of inflation, defined as
\begin{equation}
    \varepsilon \equiv \frac{\dot{H}}{H^2} \, ,
\end{equation}
we can rewrite this parameter as
\begin{equation}\label{1 inflation parameter}
    \varepsilon = \frac{\frac{3}{2}\dot{\phi}^2}{\frac{1}{2}\dot{\phi}^2 + V} \, .
\end{equation}

Inflation only occurs if $\varepsilon \ll 1$; thus, the kinetic energy density must only give small contributions to the total energy density, hence the name \textit{slow-roll inflation}. Additionally, we also must ensure that the acceleration of the scalar field is also small, allowing for the slow-roll behavior to persist. For this, the dimensionless acceleration per Hubble time is defined as 
\begin{equation}
    \delta \equiv -\frac{\Ddot{\phi}}{H \dot{\phi}} \, ,
\end{equation}
which, when it is small, ensures that the friction term is dominant and the speed of the scalar is determined by its potential, more specifically, by the slope of the potential. Moreover, ensuring a small $\delta$ is the same as ensuring that the expansion continues, as the inflation kinetic energy stays subdominant if $\delta$ is small. 

Concerning the duration of the inflationary period, it is important to note that the physical Hubble rate remains nearly constant throughout inflation. Consequently, the decrease in the comoving Hubble radius during inflation corresponds to the increase in the scale factor. This increase is typically quantified by the number of e-foldings, defined as
\begin{equation}
\vspace{-6pt}
N_{\text{tot}} \equiv \ln\left(\frac{a_{\text{end}}}{a_{\text{initial}}}\right) \, .
\end{equation}

Additionally, it is also useful to define another slow-roll parameter, $\eta$, to measure how long inflation lasts. This parameter is given by
\begin{equation}\label{accel parameter}
\vspace{-12pt}
    \eta \equiv \frac{\mathrm{d} ln(\varepsilon)}{\mathrm{d}N} = \frac{\dot{\varepsilon}}{H\varepsilon} \, ,
\end{equation}
that sometimes is also expressed and represented as
\begin{equation}
\vspace{-12pt}
    \varepsilon_{n+1} \equiv \frac{\dot{\varepsilon}_n}{H\varepsilon_n}\, ,
\end{equation}
being called the $(n + 1)$st Hubble slow-roll parameter with $n\geq 1$.  If $|\eta| < 1$, the fractional changes in $\varepsilon$ per e-fold are small, and inflation can continue. By taking the time derivatives of Equations \eqref{1 inflation parameter}  and \eqref{accel parameter}, the second slow-roll parameter can be expressed as
\begin{equation}\label{eta slow-roll}
\vspace{-6pt}
\eta = 2(\varepsilon - \delta) \, ,
\end{equation}
which shows that if $\{\varepsilon, |\delta|\} \ll 1$, then $\{\varepsilon, |\eta|\} \ll 1$ as well. 

The three parameters introduced, besides defining the regime where inflation occurs and persists, can also be used to approximate some of the previous equations resulting in new parameters, the \textit{potential slow-roll parameters}: $\varepsilon_V$ and $\eta_V$. The parameter $\varepsilon$, when much smaller than unity, allows to write Equation \eqref{Friemdann scalar} as
\begin{equation}\label{approximated FE}
    H^2 \approx \frac{V}{3 M_{Pl}^2} \, ,
\end{equation}
indicating that in this regime, the Hubble expansion is determined by the potential\linebreak of the inflaton. The approximation related with the parameter $\delta$ happens in the\linebreak Klein--Gordon equation. Considering $\delta \ll 1$, then Equation \eqref{Klein-Gordon eq} is given by
\begin{equation}\label{KG approximated}
    3H\dot{\phi} \approx -V_\phi \, ,
\end{equation}
which elucidates the relation between the slope of the potential and the \mbox{speed of inflation}. Combining the approximated Friedmann Equation \eqref{approximated FE} and the approximated \linebreak Klein--Gordon equation, the slow-roll parameter is expressed as
\begin{equation}
    \varepsilon \approx \frac{M_{Pl}^2}{2}\left(\frac{V_\phi}{V}\right)^2 \equiv \varepsilon_V \, .
\end{equation}

At last, taking the time derivative of Equation \eqref{KG approximated} and using Equation \eqref{eta slow-roll}, \linebreak  \mbox{one obtains}
\begin{equation}
    \eta \approx M_{Pl}^2 \frac{V_{\phi\phi}}{V}\equiv \eta_V \, .
\end{equation}

\subsection{Reheating: The Basics}
The rapid expansion of the universe during inflation presents a \mbox{significant challenge}, as it causes the universe to cool dramatically. To overcome this problem, the most widely accepted idea involves the inflaton field converting its energy density to a thermal bath that fills the universe at the beginning of the standard radiation-dominated epoch.  This mechanism is called \textit{reheating}, which signals the onset of the hot Big Bang phase.  Although reheating is typically viewed as a sub-topic of inflation, the complexity of its physics and the intricacies of its mechanisms make it a distinct field of research in its own right. A variety of reheating mechanisms have been explored extensively in the  literature (for a more comprehensive and detailed examination, we refer readers to the works in~\cite{Shtanov:1994ce,Felder:1999pv,Allahverdi:2010xz, Amin:2014eta,Charters:2008es}), each playing a key role in the behavior of the inflaton field.  One prominent mechanism, known as broad parametric resonance~\cite{Linde:1990flp}, involves the inflaton field oscillating around the minimum of its potential. This process leads to the production of a large number of particles, efficiently reheating the universe after around 20 oscillations. Another notable mechanism is instant preheating~\cite{Felder:1998vq}, which takes place when the inflaton field is near the bottom of its potential. During this process, particles with initially low mass are produced, but within roughly half an oscillation, these particles rapidly gain mass up to the scale associated with the Grand Unified Theory.  These massive particles then decay into heavy fermions via Yukawa interactions, which subsequently decay into lighter particles, forming a plasma composed of relativistic fluid. \mbox{As the inflaton }continues to oscillate, this plasma grows in density, eventually dominating and reheating the universe. In this review, we will cover only the fundamental aspects of classical reheating, where the energy of the inflaton is converted into particles and radiation in a thermal manner.

To understand the reheating idea, it is important to understand how inflation happens dynamically. From the dynamical point of view, inflation takes place when the inflaton field rolls slowly relative to the rapid expansion of the universe. However, as the slow-roll phase nears its end, the effective potential steepens, causing the inflaton field, denoted as $\phi$, to oscillate rapidly around the true vacuum expectation value, which is the global minimum of its potential, $\phi_0$. The coherent oscillations of the inflaton field about this global minimum, where $V(\phi_0) = V'(\phi_0) = 0$, correspond to a condensate of zero-momentum $\phi$ particles with mass $m_{\phi} = V''(\phi_0)$. These particles decay due to quantum particle creation of other fields that interact with $\phi$. The damping of the oscillations through quantum particle creation is analogous to the decay of $\phi$ particles into other lighter species to which they couple. This process can be described by the classical equation of motion for $\phi$ in the regime of rapid oscillation, which can be formulated as follows~\cite{Kolb:1990vq}:
\begin{equation}\label{reheating equation1}
\dot{\rho}_{\phi}+3H\phi+\Gamma_{\phi}\rho_{\phi}=0 \, ,
\end{equation}
with $\Gamma_{\phi}$ being the decay decay width of the $\phi$ particle. This concept of single body decay is the central point of the so-called \textit{old theory of reheating}. Additionally, the equation corresponds to the equation governing the decay of a massive particle species as a consequence to the entropy conservation. The solution for Equation \eqref{reheating equation1} is~\cite{Kolb:1990vq} 
\begin{equation}
    \rho_\phi = M^4 \left(\frac{a}{a_{osc}}\right)^{-3} e^{-\Gamma_\phi (t-t_{osc})} \, ,
\end{equation}
where $M^4$ is the vacuum energy of the scalar field at that time and $\text{osc}$ indicates the epoch when the coherent oscillations start. Ultimately, the decay products of the inflaton thermalize to a temperature, $T_{\textrm{RH}}$, given by~\cite{Kolb:1990vq}
\begin{equation}\label{reheating temperature GR1}
T_{\textrm{RH}}\simeq 0.55 g_{\ast}^{-1/4}(m_{Pl}\Gamma_{\sigma})^{1/2}\,,
\end{equation}
so that a hot thermal universe is restored.  Equation (\ref{reheating temperature GR1}) has been further developed in recent years using the Boltzmann equations as discussed in Refs.~\cite{Garcia:2020wiy,Clery:2021bwz,Garcia:2020eof,Kaneta:2022gug}.
This temperature sets an upper limit on the post-inflationary temperature and signifies the onset of the adiabatic, radiation-dominated phase of the universe. At the end of inflation, the universe can be thought of as being in a nearly ``frozen'' state. Any initial entropy would have been eliminated by inflation, leaving the energy confined to the cold, coherent motion of the inflaton field. As a result, reheating is a crucial process; any viable model of inflation must also account for how the universe was reheated and eventually thermalized to a temperature of at least around 1 MeV, in order to preserve the successful predictions \mbox{of nucleosynthesis.}

On the other, the reheating mechanism can have a deeper connection with gravity. Ford~\cite{Ford:1986sy} proposed that plasma energy primarily arises from the creation of gravitational particle production (CGPP) rather than directly from the inflaton, a mechanism known as gravitational reheating. Reheating is considered complete when the plasma constitutes the dominant energy component of the universe and while the majority of the plasma’s energy originates from the inflaton, a small portion could potentially derive from CGPP~\cite{Passaglia:2021upk}. However, this contribution is typically on the order of $H_{inf}^2 / M_{Pl}^2$, making it negligibly small. Alternatively, one could consider scenarios in which the energy from the inflaton diminishes before being transferred to a plasma. For instance, if the inflaton potential has a locally quadratic minimum, its energy density decreases slowly, behaving like matter with $\rho_\phi \propto a^{-3}$. Conversely, if the potential is sufficiently flat with $V(\phi) = 0$,\linebreak the inflaton undergoes a kination phase, with its energy density declining rapidly as $\rho_\phi \propto a^{-6}$. In the latter case, reheating would need to proceed without any \mbox{significant energy} transfer from the inflaton~\cite{Kolb:2023ydq}. These ideas and the pioneering research performed by Ford led to gravitational reheating being extensively explored by numerous studies~\cite{Copeland:2000hn,Feng:2002nb, Liddle:2003zw,Sami:2003my, Tashiro:2003qp,Chun:2009yu,Kunimitsu:2012xx, Nishizawa:2014zra,Nishi:2016wty, Dimopoulos:2018wfg,Hashiba:2018iff, Figueroa:2018twl,Haro:2018zdb, Hashiba:2018tbu,Opferkuch:2019zbd,Kamada:2019ewe,Figueroa:2024yja,Figueroa:2024asq,Yajnik:2024fjl}, especially in relation to braneworld inflation, quintessential inflation, and curvaton models. Although direct reheating through Standard Model processes, such as interactions with the Higgs, has been investigated~\cite{Figueroa:2016dsc}, most analyses involve a new, massive, and unstable particle that undergoes CGPP, mediating reheating into the Standard Model. By varying this particle’s mass and potentially its non-minimal coupling to gravity, a wide range of reheating temperatures can be achieved. Moreover, gravitational reheating is strongly\linebreak constrained~\cite{Tashiro:2003qp,Nishizawa:2014zra,Nishi:2016wty,Figueroa:2018twl} mainly due to the production of inflationary gravitational waves that develop a blue-tilted spectrum that rises toward higher frequencies.\linebreak If multiple particle species undergo CGPP, collectively carrying enough energy \mbox{to offset} the energy contained within the two polarizations of gravitational wave radiation gravitational reheating may remain feasible~\cite{Kolb:2023ydq}.

\textls[-10]{Furthermore, it should be noted that there are mechanisms in which particle production can be significantly more efficient than the scenario previously described. Specifically, when $\phi_0 = 0$, the non-adiabatic excitation of inflaton fluctuations through parametric resonance leads to a breakdown of the conventional view, where the inflaton is seen as a large collection of statistically independent particles. Instead, the spatial and temporal coherence of the inflaton can result in a dramatic shift away from the traditional theory of reheating. This phenomenon is known as \textit{preheating}~\cite{Bassett:2005xm,Charters:2005eg}. To illustrate this, consider the simple case where the inflaton is coupled to a scalar field $\chi$. The adiabatic parameter, given by~\cite{Bassett:2005xm}}
\begin{equation}
R_a \simeq \frac{m_{\phi} g^2 \phi^2}{(m_{\chi}^2 + g^2 \phi^2)^{3/2}} \,,
\end{equation}
where $g$ is a dimensionless coupling constant, indicates when the WKB approximation breaks down. The regime where $R_a \ll 1$ is referred to as the adiabatic region because, in this case, the number of particles remains an adiabatic invariant—it does not change over time, implying that no particle production occurs. Conversely, in the region where $R_a \gg 1$, the particle number is no longer conserved as an adiabatic invariant, and significant particle production is likely. In particular, non-adiabatic particle production occurs when $m_{\chi} < |g \sigma^2|$. Even if $g^2 < 0$, $R_a$ can still diverge, allowing for the production of particles with masses greater than that of the inflaton. This scenario is notable because, in cases where inflaton decay is purely perturbative, the production of such massive particles would be kinematically forbidden. For more details, see~\cite{Bassett:2005xm} and the references therein.

\subsection{Cold and Warm Inflation}

With a foundational understanding of reheating established, a profound question can be presented within the framework of the standard slow-roll inflation model:\linebreak Is it necessary to separate the expansion phase and reheating into two distinguished time periods? This question stems from two key consequences of separating these phases. First, in a cold universe, the necessary density perturbations are assumed to originate from quantum fluctuations of the inflaton field. Second, the sharp drop in temperature following the expansion necessitates a localized mechanism that can quickly convert sufficient vacuum energy to achieve reheating. Concerned by these issues, Arjun Berera and\linebreak Li-Zhi Fang proposed in~\cite{Berera:1995wh} that integrating these two distinct stages, expansion and reheating, into a single continuous process could potentially resolve the inconsistencies that arise when each is considered separately, where they demonstrated that slow-roll inflation can be parametrically consistent with the presence of a thermal component, ultimately leading to a new thermal description. This work resulted in what we today call \textit{Warm inflation}~\cite{Berera:1995ie,Berera:2023liv}. Warm inflation is a rich sub-topic of inflation, and recently one of its founders, Arjun Berera, published an exemplary review on the topic~\cite{Berera:2023liv}.

\subsection{Models for Inflation}

The fundamental physics behind inflation is not fully understood and is still being investigated. There are a multitude of models for inflation, with most coming from Effective Field theories~\cite{Cheung:2007st}. Additionally, one of the best models we currently have to describe inflation is based on modified gravity. Alexei Starobinsky~\cite{Starobinsky:1980te} observed that quantum corrections to general relativity play a significant role in understanding the early universe. These corrections generally lead to the inclusion of curvature-squared terms in the\linebreak Einstein--Hilbert action, resulting in a form of $f(R)$ modified gravity. Given the necessity of inflation in early cosmology, Starobinsky initially investigated this topic using semi-classical Einstein equations with free quantum matter fields. However, it was soon recognized that late-time inflation, relevant to the observable universe, can be primarily governed by the contribution of the squared Ricci scalar in the effective action~\cite{Starobinsky:1980te,Mukhanov:1981xt,Starobinsky:1983zz}
\begin{equation}
    S = \frac{M_{Pl}^2}{2} \int \mathrm{d}^4x \sqrt{-g}(R+\frac{R^2}{6m^2}) \, ,
\end{equation}
with $m$ being the mass of the inflaton (for an in-depth review of this model we refer to~\cite{Bamba:2015uma}). In the scalar--tensor representation, this action leads to the famous Starobinsky potential~\cite{Aldabergenov:2018qhs}
\begin{equation}
     V(\phi )=\frac{3}{4}M_{Pl}^2m^2\left(1-e^{-{\sqrt {2/3}}\phi /M_{Pl}}\right)^{2}\, .
\end{equation}

Modified gravity presents a natural ability to tackle the theoretical origin of inflation with a plethora of models, in addition to the aforementioned model. For this subject, we refer to~\cite{Nojiri:2017ncd,Odintsov:2023weg,Myrzakulov:2013hca,Clifton:2011jh} for an in-depth review of such models and also~\cite{Gamonal:2020itt,Jamil:2013nca,HosseiniMansoori:2023zop,Feng:2022vcx,Sadatian:2024inx} for models that are not contained in the previous references. Furthermore, inflation can also have a connection with high-dimension theories such as Branes~\cite{Dvali:2001fw,Garcia-Bellido:2001lbk,Kachru:2003sx,HenryTye:2006uv}, Strings~\cite{Baumann:2014nda,Stewart:1994ts,Baumann:2009ni}, and Supergravity~\cite{Stewart:1994ts,Kallosh:2013yoa,Ferrara:2013rsa,Kallosh:2017ced}. The standard model with the Higgs field also shows a possible role in inflation~\cite{Bezrukov:2007ep,Rubio:2018ogq,Bezrukov:2011gp}. There is a wide variety of models for inflation with all the various origins. In Table~\ref{tab:1}, we present the models that agree with the Planck data~\cite{Planck:2018vyg}.

\begin{table}[H]
\caption{A selection of slow-roll inflationary models that are in agreement with the Planck data~\cite{Planck:2018vyg}. The models were chosen based on the criteria presented in {Table 5 }of~\cite{Planck:2018vyg} with the Starobinsky model as reference.}
\label{tab:1}
\begin{adjustwidth}{-\extralength}{0cm}
		\newcolumntype{C}{>{\centering\arraybackslash}X}
		\begin{tabularx}{\fulllength}{CCC}
			\toprule
\textbf{Model} & \textbf{ Potential }& \textbf{ Parameter Range} \\ 
\midrule
Starobinsky &
  $\Lambda^4 \left( 1 - e^{- \sqrt{2/3} \phi/M_\mathrm{Pl} } \right)^2$ &
  \dots \\ 
Power-law~\cite{McAllister:2008hb}&
  $\lambda M_\mathrm{Pl}^{3} \phi$ &
  \dots \\ 
Power-law~\cite{Silverstein:2008sg}&
  $\lambda M_\mathrm{Pl}^{10/3} \phi^{2/3}$ &
 \dots \\ 
Non-minimal coupling~\cite{Garcia-Bellido:2008ycs} &
  $\lambda^4 \phi^4 + \xi \phi^2 R/2 $ &
  $0.3 < \log_{10} (\mu_2 /M_\mathrm{Pl}) < 4.85$ \\
Hilltop quadratic model~\cite{Boubekeur:2005zm} &
  $\Lambda^4 \left( 1 - \phi^2 / \mu_2^2 + \dots \right)$ &
  $0.3 < \log_{10} (\mu_2 /M_\mathrm{Pl}) < 4.85$ \\ 
Hilltop quartic model~\cite{Boubekeur:2005zm}&
  $\Lambda^4 \left( 1 - \phi^4 / \mu_4^4 + \dots \right)$ &
  $-2 < \log_{10} (\mu_4 /M_\mathrm{Pl}) < 2$ \\ 
D-brane inflation $(p = 2)$~\cite{Dvali:2001fw,Garcia-Bellido:2001lbk,Kachru:2003sx}&
  $\Lambda^4 \left( 1 - \mu^2_{{\rm D} \, 2} / \phi^p + \dots \right)$ &
  $- 6  < \log_{10} (\mu_{{\rm D} \, 2} /M_\mathrm{Pl}) < 0.3$ \\ 
D-brane inflation $(p = 4)$~\cite{Dvali:2001fw,Garcia-Bellido:2001lbk,Kachru:2003sx}&
  $\Lambda^4 \left( 1 - \mu^4_{{\rm D} \, 4} / \phi^p + \dots \right)$ &
  $- 6  < \log_{10} (\mu_{{\rm D} \, 4} /M_\mathrm{Pl}) < 0.3$ \\
Potential with exponential tails~\cite{Goncharov:1984jlb,Cicoli:2008gp}&
  $\Lambda^4 \left[ 1 - \exp{\left(-q \phi / M_\mathrm{Pl}\right)} + \dots \right]$ &
  $- 3  < \log_{10} q < 3$ \\ 
E-model $(n = 1)$~\cite{Kallosh:2013yoa,Ferrara:2013rsa}& $\Lambda^4 \left\{ 1 - \exp{\left[- \sqrt{2} \phi \left(\sqrt{3 \alpha_1^{\rm E}} M_\mathrm{Pl}\right)^{-1}\right] }\right\}^{2 n} $
   &  $-2 < \log_{10} \alpha_1^{\rm E} < 4$
   \\ 
E-model $(n = 2)$~\cite{Kallosh:2013yoa,Ferrara:2013rsa}&  $\Lambda^4 \left\{ 1 - \exp{\left[- \sqrt{2} \phi \left(\sqrt{3 \alpha_2^{\rm E}} M_\mathrm{Pl}\right)^{-1}\right] }\right\}^{2 n} $
   & $-2 < \log_{10} \alpha_2^{\rm E} < 4$
   \\ 
T-model $(m = 1)$~\cite{Kallosh:2013yoa,Ferrara:2013rsa} & $\Lambda^4 \tanh^{2 m} \left[ \phi \left(\sqrt{6 \alpha_1^{\rm T}} M_\mathrm{Pl}\right)^{-1}\right] $
   & $-2 < \log_{10} \alpha_1^{\rm T} < 4$
   \\ 
T-model $(m = 2)$~\cite{Kallosh:2013yoa,Ferrara:2013rsa} &  $\Lambda^4 \tanh^{2 m} \left[ \phi \left(\sqrt{6 \alpha_2^{\rm T}} M_\mathrm{Pl}\right)^{-1}\right] $
   & $-2 < \log_{10} \alpha_2^{\rm T} < 4$
   \\ 
\bottomrule
		\end{tabularx}
	\end{adjustwidth}
\end{table}

\section{Brief Thermal History of the Universe}\label{sec:Brief Thermal History of the Universe}

With all the previous subjects explored, we can now provide a brief description of the thermal history of the universe, as most of its story can be constructed with the notions previously presented. With this being said, the thermal history of the universe is roughly the following:

\begin{enumerate}
\item \textbf{\emph{Quantum Gravity?}} ($T>T_{Pl}\sim 10^{19}$ GeV
) Below this point, quantum corrections to GR should make it invalid, and a theory of quantum gravity is expected to be necessary to obtain a correct description. In inflationary scenarios, this is usually called pre-inflationary cosmology.  
\item \textbf{\emph{Inflation}}. Epoch of accelerated expansion of the universe, which is likely exponential in nature, can be characterized by the de Sitter solution. This solution is defined\linebreak by Equation \eqref{deSitterGR}. During this period in the universe's history, adiabaticity breaks down, and equilibrium thermodynamics does not hold.
\item \textbf{\emph{End of Inflation and particle production}}. In this period, dark expanding ``emptiness'' filled by the scalar (or some other) field, inflaton, exploded, creating light and other elementary particles. This is the period of reheating. 
\item  \textbf{\emph{Beginning of the radiation dominated universe}}. The universe becomes\mbox{ well described} by equilibrium thermodynamics, being adiabatically cooled down. If current ideas are correct, during this epoch, the universe underwent several phase transitions.\linebreak During the stages at which phase transitions occur, adiabaticity could be broken. 
\item \textbf{\emph{Grand unification phase transition}} ($T\sim 10^{15}$ GeV--$10^{17}$ GeV)~\cite{Kolb:1990vq}. The objective of Grand Unified Theories (GUTs) is to unify the electromagnetic, electroweak, and strong interactions into a single gauge group that should remain a valid symmetry at the highest energy scales. As the energy decreases, the theory undergoes a series of spontaneous symmetry breakings (SSBs) into successive subgroups. Cosmologically, these SSBs would correspond to phase transitions occurring during the evolution of the universe, potentially leading to the formation of topological defects (see, for instance, Ref.~\cite{ParticleDataGroup:2012pjm}). Specifically, at a certain point, the strong force separates from the other fundamental forces. This occurs as the temperature drops below a threshold where supermassive gauge bosons X and Y, as well as supermassive Higgs (H) bosons, can be created. As the temperature continues to decrease, these supermassive bosons (X, Y, and H) eventually decay, possibly violating baryon number conservation and/or generating entropy. It should be emphasized, however, that in inflationary scenarios, this phase of the thermal history can be situated before the period of inflation.
\item \textbf{\emph{EW phase transition}} ($T\sim 140$ GeV)~\cite{Ramsey-Musolf:2019lsf}. Some of the gauge bosons and other particles acquire mass via the Higgs mechanism. 
\item \textbf{\emph{QCD phase transition}} ($T\sim 155$ MeV)~\cite{Guenther:2020jwe}. Quarks lose their so-called \emph{asymptotic freedom}, a property they exhibit at high energies, resulting in the absence of free quarks and gluons. Consequently, the \emph{quark--gluon} plasma transitions into a \emph{hadron} gas. During this process, composites of three quarks (baryons) and quark--antiquark pairs (mesons) are formed. 
\item \textbf{\emph{Neutrino decoupling}} ($T\sim 1$ MeV)~\cite{Akita:2022hlx}. Prior to this point in the history of the\linebreak universe, neutrinos were maintained in thermal equilibrium through weak\linebreak interactions of the type $\bar{\nu}\nu \leftrightarrow e^{+}e^{-}$, where $\nu$ and $\bar{\nu}$ represent generic neutrinos and\linebreak antineutrinos, respectively, and $e^{+}$ and $e^{-}$ represent positrons and electrons,\linebreak respectively. The reaction rate for such processes is approximately given by~\cite{Kolb:1990vq}
\begin{equation}
\Gamma_{int}\simeq G_F^2 T^5\,,
\end{equation}
where $G_F\simeq 1.166 \times 10^{-5}\, {\rm GeV^{-2}}$ is the Fermi constant. Comparison of this rate with the expansion rate of the universe, 
\begin{equation}
\frac{\Gamma_{int}}{H}\simeq \frac{G_F^2T^5 m_{Pl}}{T^2}\simeq \left(\frac{T}{1~\textrm{MeV}}\right)\,,
\end{equation}
leads to a freezing temperature of about $1$ MeV. Before neutrino decoupling, it is generally assumed that all relativistic species are in thermal equilibrium with the photons. These species include the photons themselves, with $g_{\gamma} = 2$ degrees of freedom; the electrons and positrons, which together have $g_{e^{\pm}} = 4$ degrees of freedom; and the nearly massless neutrinos, with $g_{\nu} = 2N_{\nu}$ degrees of freedom, where $N_{\nu}$ is the number of generations of nearly massless neutrinos. Therefore, before neutrino decoupling, the effective number of degrees of freedom for entropy, $g_{\ast S}$, is given by $2 + 7(2 + N_{\nu})/4$. After decoupling, the neutrinos, once in thermal contact with the plasma, no longer share the same temperature with the photons due to the weak interactions being insufficient to maintain equilibrium during subsequent stages. Consequently, from this point onward, $g_{\ast S}$ should be calculated using $\left[22 + 7N_{\nu}(T_{\nu}/T_{\gamma})^3\right]/4$.
\item \textbf{\emph{Electron--positron annihilation}} ($T\sim 0.5$ MeV). Shortly after the neutrino decoupling, the temperature drops below the mass of the electron--positron, which thereby become non-relativistic. Thereafter, the entropy in the $e^{+}e^{-}$ is transferred to the photons but not to the neutrinos, which are already decoupled from the thermal plasma.\linebreak Therefore,  conservation of the entropy per comoving volume implies that
\begin{eqnarray}
\frac{S_{\textrm{before annihilations}}}{S_{\textrm{after annihilations}}} &=& \frac{S_{\textrm{at the decoupling}}}{S_{\textrm{after annihilations}}}
	\nonumber \\
&=&\left(\frac{22+7N_{\nu}}{4}\right)\,\left(\frac{a_{d_{\nu}}T_{d_{\nu}}}{a T_{\gamma}}\right)^3\Big/\left[2+\frac{7}{4}N_{\nu}\left(\frac{T_{\nu}}{T_{\gamma}}\right)^3\right]
\nonumber \\
&=&1,\label{after electron annihilation}
\end{eqnarray}
where $T_{\nu}$ is the characteristic temperature of a generic neutrino, and $T_{d_{\nu}}$ and $a_{d_{\nu}}$ represent, respectively, the temperature of the generic neutrino  and the value of the scale factor at the moment of the decoupling and, since the effective temperature of the neutrino behaves after the decoupling accordingly with Equation~\eqref{temperature of massless decoupled species}, that is $T_{\nu}=T_{d_{\nu}} a_{d_{\nu}}/a$, it is straightforward to conclude from Equation~\eqref{after electron annihilation} that $T_{\nu}/T_{\gamma}=(4/11)^{1/3}$. At the present time, the species known to be relativistic are the neutrinos and the photons, and thereby, it can be easily established that the effective degrees of freedom in the energy density in radiation and in the entropy density are presently given by
\begin{eqnarray}
g_{{\ast,0}}&=&2+\frac{7}{4}\,\left(\frac{4}{11}\right)^{4/3}N_{\nu}=2+0.454N_{\nu}\,,
\label{present day degrees of freedom in radiation}
	\\
g_{\ast S,0}&=&2+\frac{7}{11}N_{\nu}=2+0.636N_{\nu}\,.
\label{present day degrees of freedom in entropy}
\end{eqnarray}

It becomes obvious that $g_{\ast S}(T_{\gamma})\neq g_{\ast}(T_{\gamma})$. This is due to the fact that at the present time, both the photons and the neutrinos are decoupled, and therefore their entropies are separately conserved, a fact that is used implicitly when considering $T_{\gamma}\neq T_{\nu}$. Yet, it must be emphasized that for most of the history of the universe subsequent to the inflationary epoch, all relativistic particle species share the same temperature, and thus, $g_{\ast S}$ can be replaced in most situations by $g_{\ast}$ as already pointed out.
\item \textbf{\emph{Big Bang Nucleosynthesis}} ($T\simeq 10-0.1$ MeV
). At about $1$ MeV, the ratio between the neutron number and the proton number (usually termed \emph{neutron-to-proton} ratio) freezes out. Shortly thereafter ($T\sim 0.1$ MeV), the synthesis of light elements begins (for more in-depth studies, see~\cite{Cyburt:2015mya,Iocco:2008va,Fields:2019pfx}). 
\item  \textbf{\emph{Matter--Radiation Equality}}. 
After nucleosynthesis, the universe reaches a point after which the matter comes to dominate over the radiation. According to the stages of the universe previously defined, such a point marks the entrance of the universe in its adolescence, and it is usually called the \emph{matter--radiation equality}. So, comparing the density profile for radiation with the density profile of matter in the single fluid interpretation and defining $\rho_M$ as the total energy density in matter, that is, the energy density which is not in the form of radiation or some sort of vacuum energy, it is also possible to find that the matter--radiation equality is attained at a redshift, $z_{eq}$,\linebreak given by
\begin{equation}
1+z_{eq}=\frac{\rho_{M,0}}{\rho_{R,0}}=\frac{30}{\pi^2}\frac{\Omega_{M,0} \rho_{c,0}}{g_{\ast S,0}T_{\gamma,0}^4}\,,\label{matter equals radiation 1}
\end{equation}
where in the last step, the equation $\rho_{R,0}=\pi^2 g_{\ast,0}/30T_{\gamma,0}$ is used in order to establish the present day radiation energy density. On the other hand, noticing that since the annihilation of the electrons $\mathcal{S}_{\ast}$ is effectively constant and given by $g_{\ast S,0}$, it is possible from the relation between the red shift and the cooling of the universe provided by Equation~\eqref{redshift  and cooling characterization} to establish that the temperature, $T_{eq}$, at which matter equals radiation, is given by
\begin{equation}
T_{eq}=\frac{30}{\pi^2}\frac{\Omega_{M,0}\rho_{c,0}}{g_{\ast S,0}T_{\gamma,0}^3}\,.
\end{equation}

This turning point marks the onset of structure formation.  
\item \textbf{\emph{Photon decoupling and recombination}} ($T\sim 0.2$ MeV--$0.3$ MeV). In the early universe, the interactions between photons and electrons were rapid when in comparison to the expansion rate of the universe, and so radiation (photons) and matter (electrons, protons and nuclei) were kept in good thermal contact, thus remaining in equilibrium. However, eventually, the universe reached a point were the thermal contact between these two components was no longer maintained and radiation decoupled from matter. The threshold temperature marking this point can be computed by means of the usual procedure ($\Gamma_{\gamma}=H$) and the appropriate reaction rate, which for this case is given by the electron scattering rate
\begin{equation}
\Gamma_{\gamma}=n_{e}\sigma_T \,,\label{thomson scattering reaction rate 1}
\end{equation}
where $n_e$ is the number density of free electrons and $\sigma_T=8\pi \alpha ^2_{em}/(3m_e^2)=6.65\times 10^{-25}$ cm$^{2}$ is the Thomson cross section. Note that indeed this is the relevant reaction because if on the one hand the scattering from protons is irrelevant, since its corresponding cross section is suppressed by a factor $1/m_p^2$, on the other hand the electrons bound into hydrogen atoms have their charge effectively shielded and thus do not contribute to Thomson scattering. From the reaction rate given by (\ref{thomson scattering reaction rate 1}), it can be seen that for as long as the free electrons are abundant, radiation and matter are kept in thermal equilibrium; however, as temperature decreases, electrons and nuclei start to combine to form neutral atoms. This is the period of \emph{recombination} and corresponds to an epoch of the density number of free electrons sharply falling. During this period, the photon mean free path $\lambda\simeq \Gamma_{\gamma}^{-1}$ grew rapidly or, if preferred, $\Gamma_{\gamma}$ decreased rapidly, and soon a point was reached where it finally became longer than the horizon distance ($H^{-1}$). This point marked the decoupling of photons from matter, as their interactions with each other were no longer able to maintain them in thermal equilibrium, and from this point on, the universe became transparent to radiation. Today, these photons manifest themselves as the CMB, and after this point, $T$ refers to the photon temperature $T_{\gamma}$. The observed CMB temperature today is~\cite{Planck:2018vyg,Fixsen:2009ug}
\begin{equation}
    T_{\gamma,0}=2.7255\pm0.0006~\textrm{K} \, .
\end{equation}
\item \textbf{\emph{The Dark Age}}. There were no stars yet, and as the cosmic expansion redshifted CMB photons towards the infrared, they became invisible heat radiation and became completely dark. This lasted for several hundred million years.  
\item \textbf{\emph{Formation of the first stars and reionization}}. In the darkness, the seeds of structure formation were already planted, and as masses began to gather, the first stars lit up one by one.
\item \textbf{\emph{Present time $t_0$}}.
\end{enumerate}

\section{Summary and Conclusions}\label{Sec:Conclusion}

The primordial epoch of the universe is a period of extraordinary importance for understanding the broader context of cosmological evolution. This phase, which occurred shortly after the Big Bang, laid the groundwork for the formation of all structures we observe in the universe today, from galaxies to stars and planets. In this review, we delved into the thermodynamic framework that governs this early stage, drawing on insights from both the Standard Model of cosmology and the Standard Model of particle physics.\linebreak These two models, which are among the most well-established theories in their respective fields, provide a comprehensive description of the physical laws and interactions that shaped the universe during its infancy.
For most of the early universe, a very good\linebreak approximation is that it was in thermal equilibrium. This means that, during this time, \linebreak the particles and radiation were uniformly distributed and in balance, allowing\linebreak the system to be effectively described using the principles of equilibrium thermodynamics.\linebreak By applying this framework in conjunction with the Cosmological Principle, which states that the universe is homogeneous and isotropic on large scales, we can define fundamental quantities such as number density, energy density, and pressure. These quantities are essential for characterizing the state of the universe at different points in its evolution.

Given the complexity of calculating these quantities, we distinguish between two key regimes: the relativistic regime, where the temperature ($T$) is much greater than the particle mass ($m$), and the non-relativistic regime, where the temperature is much lower than the particle mass. These two regimes allow us to categorize the primordial universe into distinct phases, each with its own thermodynamic properties. The non-relativistic phase, in particular, plays a crucial role in constructing the timeline of the universe’s evolution, as it marks the period where particles began to cool and form the structures that would eventually lead to galaxies and stars.
Moreover, the second law of thermodynamics provides a vital link between entropy and these fundamental quantities. In the early universe, the total entropy density was, to a very good approximation, primarily determined by the contributions from relativistic species. The conservation of entropy density, which is a key principle in thermodynamics, allows us to compute the asymmetry between matter and antimatter. This asymmetry is one of the most profound mysteries of the primordial universe, as it ultimately led to the dominance of matter over antimatter, enabling the existence of the matter-based universe we observe today.

While the early universe is often considered to have been in thermal equilibrium, maintaining such an equilibrium in an expanding universe is a complex challenge that requires careful consideration. As the universe expands, its temperature decreases, which affects the ability of particles to interact and remain in equilibrium. Thus, one of the key discussions explored in this work is the point at which the universe's temperature drops low enough for decoupling events to occur. Decoupling refers to the moment when certain particles no longer interact frequently enough to remain in thermal equilibrium with the rest of the universe. When the temperature decreases to a certain threshold, particles in the non-relativistic regime, where their thermal energy is much lower than their rest mass, begin to decouple. This shift marks the transition to a non-relativistic scenario, where the dynamics of the universe are primarily governed by these decoupled particles.
The analysis of this transition is carried out using the concept of effective degrees of freedom, which represents the number of ways in which particles can store energy. This concept is applied to each major thermodynamic quantity, such as energy density and entropy, to understand how the behavior of the universe changes as it cools and expands. The reduction in effective degrees of freedom during decoupling is a crucial factor in shaping the subsequent evolution of the universe.

We also examine the critical role of out-of-equilibrium phenomena, which are\linebreak essential for understanding systems that have moved beyond thermal equilibrium.\linebreak A key tool in this analysis is the Boltzmann equation, a fundamental equation in\linebreak statistical mechanics that models the behavior of particle distributions over time.\linebreak Using the Boltzmann equation, we develop a statistical framework that leads to two significant results: the freezing of ``hot relics'' and the freezing of ``cold relics''.
``Hot relics'' are particles that decouple from the thermal bath while still in the relativistic regime, meaning they retain substantial kinetic energy even after decoupling. In contrast, ``cold relics'' decouple during the non-relativistic phase, when their speeds are much lower.\linebreak The freezing of these relics is a pivotal event in the universe's thermal history, as it determines the distribution and abundance of particles that would eventually become the building blocks of galaxies, stars, and other cosmic structures.
By integrating insights from both equilibrium and out-of-equilibrium processes, we construct a comprehensive timeline of the universe's evolution. This timeline outlines the major epochs and events, from the earliest moments after the Big Bang to the formation of the first atoms, galaxies, and beyond. Each phase of the universe's history is briefly described, highlighting how the interplay between thermal equilibrium and out-of-equilibrium phenomena has shaped the cosmos into its present form.

In conclusion, mastering the theoretical framework of thermodynamics during the primordial epoch of the universe is crucial for comprehensively understanding the sequence of pivotal events that shaped the cosmos before this era concluded.  This framework provides the foundation for analyzing the intricate processes that occurred in the earliest moments after the Big Bang, offering insights into how the universe evolved into its current state.
One of the first major events in this timeline is cosmic inflation, a rapid expansion that occurred fractions of a second after the Big Bang. Understanding the thermodynamic conditions during inflation is essential for explaining how the universe transitioned from a chaotic, high-energy state to a more stable one. This transition set the stage for the formation of the large-scale structure of the universe and helped address several fundamental issues in cosmology, such as the horizon and flatness problems.
Another critical phenomenon is the generation of baryonic asymmetry as mentioned above, which refers to the observed imbalance between matter and antimatter in the universe. Theoretical models suggest that certain processes during the early universe, likely occurring out of thermal equilibrium, created this asymmetry. By applying thermodynamic principles to these scenarios, we can explore the possible mechanisms that led to the dominance of matter over antimatter, a key factor that allowed galaxies, stars, and planets to form.

Following these events, the process of nucleosynthesis, where the first atomic nuclei were formed, marks another significant milestone. This process, which occurred within the first few minutes after the Big Bang, is governed by the interplay between particle physics and thermodynamics. Understanding the temperature and density conditions during this period allows us to predict the relative abundances of the light elements, such as hydrogen, helium, and lithium, that we observe today. These predictions have been remarkably successful, providing strong evidence for the Big Bang model.
As the universe continued to expand and cool, the formation of the cosmic microwave background (CMB) became a defining moment. The CMB is the relic radiation from the time when the universe became transparent, approximately 380,000 years after the Big Bang. The thermal description of the universe during this period is essential for understanding the properties of the CMB, which carries a wealth of information about the early universe, including the seeds of \mbox{cosmic structure.}

Large-scale matter creation is essential in cosmology, as it explains the origin of\linebreak large-scale structures and the evolution of cosmological fluid components. In this context, the study of matter creation within the framework of irreversible thermodynamics of open systems is particularly relevant. The investigation of particle production in expanding universes began with Erwin Schrödinger’s pioneering work in 1939–1940~\cite{Schrodinger:1939, Schrodinger:1940}, where he observed that a scalar particle could spontaneously produce pairs of particles. Leonard Parker later extended this idea, showing that photons could stimulate the creation of pairs of photons in an expanding FLRW universe~\cite{Parker:2012at}. Despite Schrödinger’s findings, early research lacked the mathematical tools to fully support these ideas, and it was not until the late 1960s that Parker formalized a mechanism for gravity-induced particle creation using quantum field theory in curved spacetime~\cite{Parker:1968mv, Parker:1969au, Parker:1971pt, Parker:1972kp}. His work linked FLRW geometry to particle production, with implications for inflationary perturbations~\cite{Parker:2012at}.

In the 1980s, Ilya Prigogine and collaborators~\cite{Prigogine:1988jax, Prigogine:1986, Prigogine:1989zz} proposed an alternative\linebreak cosmological model based on irreversible thermodynamics to reconcile particle creation with entropy increase. They argued that the adiabatic, reversible nature of the Einstein field equations could not explain irreversible particle creation. Their model introduced\linebreak an irreversible creation term in the energy--momentum tensor. However, this framework lacked a clear physical interpretation within GR, where energy conservation posed limitations. More recently, modified gravity theories that include non-minimal geometry–matter couplings~\cite{Bertolami:2007gv,Bertolami:2008ab,Harko:2010mv,Harko:2011kv,Haghani:2013oma,Harko:2014sja,Harko:2014aja,Harko:2014gwa,Barrientos:2018cnx,Harko:2018gxr} have provided new insights into Prigogine’s approach.\linebreak In these theories, non-conservation of the energy--momentum tensor allows for a physical interpretation of irreversible energy flow from the gravitational field to matter, leading to particle creation~\cite{Harko:2014pqa}. The effects of this process on cosmological evolution have been studied in various models~\cite{Harko:2014pqa, Pinto:2022tlu, Harko:2015pma, Cipriano:2023yhv, Pinto:2023phl}.
In summary, while several mechanisms can\linebreak generate particles, the framework of non-minimal geometry--matter couplings, where the\linebreak energy--momentum tensor is not conserved, offers a natural approach to studying particle creation through the thermodynamics of open systems~\cite{Harko:2014pqa}.

In the present review, we have focused on the thermodynamics of the phenomena taking place during the main stages of the primordial universe, within the framework of the Standard Cosmological Model.
A more ambitious and exhaustive review should also envisage  the investigations  carried out on a substantially wider list of questions and goals regarding the thermodynamics  of gravitational and cosmological physics.  For instance, it has been well known since the 1960s that black holes allow a remarkable  thermodynamical interpretation~\cite{Bekenstein:1973ur,Bekenstein:1974ax,Bardeen:1973gs,Hawking:1976de}. This has been the object of  a multitude of  relevant studies~\cite{Carlip:1999db,Wald:1997qp,Gibbons:1977mu,Visser:1992qh,Hayward:1993wb,Peca:1998cs,Peca:1998dv,Lemos:2013dsa,Andre:2019zzo}, and for a comprehensive  survey of this topic, the reader is invited to consult~\cite{Wald:1999vt} and the references therein. The thermal physics of other exotic gravitational objects, namely, the case of wormholes has also attracted a considerable interest~\cite{Gonzalez-Diaz:1995mzx,Gonzalez-Diaz:2004clm,Martin-Moruno:2009yni,Bandyopadhyay:2012us,Debnath:2012fqp,Rehman:2020myc}. More generally, the thermodynamics of the expanding universe has been investigated from several viewpoints~\cite{Abreu:2010sc,Abreu:2011fr,Santiago:2018lcy,Santiago:2018kds,Mimoso:2013zhp,Mimoso:2016jwg,Pavon:1990ha,Quevedo:1994yg,Triginer:1995kx,Zimdahl:1996tg,Zimdahl:2000hq,delCampo:2011aa,Pavon:2012qn,Gonzalez-Espinoza:2019vcy,Banerjee:2023rvg,Banerjee:2023evd,Lima:1990cj,Lima:1995kx,Lima:1995kt,Carrillo:1996yu,Coley:1995dpj,Maartens:1995wt,Gunzig:1997tk,Maartens:1997sh,Herrera:1989zza,Herrera:1997xq,Herrera:1997nj,Sussman:1999vx,Herrera:2001ff,Herrera:2004xc,Herrera:2004ip,Herrera:2006qvi,Herrera:2020gdg,bak,gmi,bw,Cai:2006rs,setare1,yg,ninfa,karami,saha,Karami:2010zz,Mazumder:2009zz,Chakraborty:2010jx,Chakraborty:2012cw,Pereira:2008af,Lima:2015mca,Komatsu:2021ncs,Cafaro:2021vhy,Sanchez:2022xfh,Komatsu:2023wml,Padmanabhan:2009vy,Padmanabhan:2013nxa,Padmanabhan:2013xyr,Padmanabhan:2015zmr}. 
In particular, some geometrically motivated approaches have been devised~\cite{Penrose:1994de,Binetruy:2014ela,Callan:1994py,Bousso:1999xy,Jacobson:1995ab,Espinosa-Portales:2021cac,Arjona:2021uxs,Espinosa-Portales:2022jii,Garcia-Bellido:2024qau,Azevedo:2019oah,Avelino:2020fek,Avelino:2015kdn}. 
The latter topics, albeit fascinating, fall beyond the scope of the present review, and will be addressed in a follow-up work.

Thus, summing up, by integrating the thermodynamic framework with the two most robust models in modern science, the standard model of cosmology and the standard model of particle physics, we gain a powerful tool for making precise predictions.\linebreak This combination allows us to construct a coherent narrative of the universe's early history, from the chaotic moments following the Big Bang to the formation of the first galaxies.\linebreak It also provides a foundation for exploring new physics beyond the standard models, as any deviation from predicted outcomes can point to areas where our current understanding might need refinement.
In essence, the thermodynamics of the primordial universe is not just a theoretical exercise; it is a cornerstone of modern cosmology that enables us to connect the dots between fundamental physics and the observable universe. By continuing to refine this framework, we can deepen our understanding of the universe's origins and uncover new insights into the forces that have shaped its evolution.

\vspace{6pt}



\authorcontributions{Conceptualization, D.S.P. and J.F.; methodology, D.S.P., J.F., F.S.N.L. and J.P.M.; software, D.S.P. and J.F.; validation, D.S.P., J.F., F.S.N.L. and J.P.M.; formal analysis, D.S.P., J.F., F.S.N.L. and J.P.M.; investigation, D.S.P., J.F., F.S.N.L. and J.P.M.; resources, D.S.P., J.F., F.S.N.L. and J.P.M.; writing---original draft preparation, D.S.P. and J.F.; writing---review and editing, D.S.P., J.F., F.S.N.L. and J.P.M.; visualization, D.S.P., J.F., F.S.N.L. and J.P.M.; supervision, F.S.N.L. and J.P.M.; project administration, F.S.N.L. and J.P.M.; funding acquisition, F.S.N.L. and J.P.M. All authors have read and agreed to the published version of the manuscript.}

\funding{ This research was funded by the Fundação para a Ciência e a Tecnologia (FCT) from the research grants UIDB/04434/2020, UIDP/04434/2020, and PTDC/FIS-AST/0054/2021.}

\institutionalreview{Not applicable.}

\informedconsent{Not applicable.}

\dataavailability{Data are contained within the article.}

\acknowledgments{F.S.N.L. also acknowledges support from the Fundação para a Ciência e a Tecnologia (FCT) Scientific Employment Stimulus contract with reference CEECINST/00032/2018.}

\conflictsofinterest{The authors declare no conflicts of interest.}

\clearpage 
\abbreviations{Abbreviations}{
The following abbreviations are used in this manuscript:\\

\noindent 
\begin{tabular}{@{}ll}
SMC & Standard Model of Cosmology\\
FLRW & Friedmann--Lema{\i}tre--Robertson--Walker \\
CMB & Cosmic Microwave Background\\
SM & Standard Model of Particles\\
MB & Maxwell--Boltzmann\\
FD & Fermi--Dirac \\
BE & Bose--Einstein \\
\end{tabular}
}
\begin{adjustwidth}{-\extralength}{0cm}

\reftitle{References}
\PublishersNote{}
\end{adjustwidth}
\end{document}